\documentclass[aps,prb,twocolumn,superscriptaddress]{revtex4}

\usepackage{graphicx,bm,amssymb,amsmath,xcolor,pifont,soul }
\usepackage[linktocpage=true,colorlinks=true,pdfborder={0 0 0},linkcolor=blue,citecolor=red,filecolor=yellow,urlcolor=blue,bookmarks,pdfauthor={},]{hyperref}
\usepackage[normalem]{ulem}
\usepackage{epstopdf}
\usepackage{epsfig}

\def\ve{{\varepsilon}}
\def\w{\omega}
\def\bk{{\bf k}}
\def\bq{{\bf q}}
\def\bkpq{{\bf kpq}}

\def\>{\rangle}
\def\<{\langle}

\def\sq3{\sqrt{3} \times \sqrt{3} \times 1}

\def\s1{1 \times 1 \times 1}
\def\inv3{(1/3, 1/3, 0)}
\def\c2{2 \times 2 \times 2}

\def\intbz{\int\frac{d\bk}{\Omega_{\rm {BZ}}}}

\def\pqvna{\hbar\Pi_{\bq\nu}^{\rm {NA}}(\omega)}

\def\pqv{\Pi_{\bq\nu}}
\def\summn{\sum\limits_{mn}}
\def\gmnvkqb{g_{mn,\nu}^b(\bk, \bq)}
\def\gmnvkqs{g_{mn,\nu}^*(\bk, \bq)}
\def\gmnvkq{g_{mn,\nu}(\bk, \bq)}

\def\fnk{f_{n\bk}}
\def\fmkq{f_{m\bkpq}}
\def\enk{\ve_{n\bk}}
\def\emkq{\ve_{m\bkpq}}
\def\bkpq{{\bf k}+{\bf q}}
\def\psimkq{\psi_{m\bkpq}}
\def\psink{\psi_{n\bk}}
\def\pv{\partial_{\bq\nu}V}
\def\pqvnaw{\Pi_{\bq\nu}^{\rm {NA}}(\omega)}
\def\pqva{\Pi_{\bq\nu}^{\rm A}}

\begin{document}

\title{Electronic, vibrational, and electron-phonon coupling properties in SnSe$_2$ and SnS$_2$ under pressure}


\author{Gyanu Prasad Kafle}
\affiliation{Department of Physics, Applied Physics, and Astronomy, Binghamton University-SUNY, Binghamton, New York 13902, USA}
\author{Christoph Heil}
\affiliation{Institute of Theoretical and Computational Physics, Graz University of Technology, NAWI Graz, 8010 Graz, Austria}
\author{Hari Paudyal}
\affiliation{Department of Physics, Applied Physics, and Astronomy, Binghamton University-SUNY, Binghamton, New York 13902, USA}
\author{Elena R. Margine}
\email{rmargine@binghamton.edu}
\affiliation{Department of Physics, Applied Physics, and Astronomy, Binghamton University-SUNY, Binghamton, New York 13902, USA}
\date{\today}

\begin{abstract}
The tin-selenide and tin-sulfide classes of materials undergo multiple structural transitions under high pressure leading to periodic lattice distortions, superconductivity, and topologically non-trivial phases, yet a number of controversies exist regarding the structural transformations in these systems. 
We perform first-principles calculations within the framework of density functional theory and a careful comparison of our results with available experiments on SnSe$_2$ reveals that the apparent contradictions among high-pressure results can be attributed to differences in experimental conditions. 
We further demonstrate that under hydrostatic pressure a $\sq3$ superstructure can be stabilized above 20~GPa in SnS$_2$ via a periodic lattice distortion as found recently in the case of SnSe$_2$, and that this pressure-induced phase transition is due to the combined effect of Fermi surface nesting and electron-phonon coupling at a momentum wave vector $\bq$ = $\inv3$.
In addition, we investigate the contribution of nonadiabatic corrections on the calculated phonon frequencies, and show that the quantitative agreement between theory and experiment for the high-energy $A_{1g}$ phonon mode is improved when these effects are taken into account. 
Finally, we examine the nature of the superconducting state recently observed in SnSe$_2$ under nonhydrostatic pressure and predict the emergence of superconductivity with a comparable critical temperature in SnS$_2$ under similar experimental conditions. Interestingly, in the periodic lattice distorted phases, the critical temperature is found to be reduced by an order of magnitude due to the restructuring of the Fermi surface. 

 \end{abstract}	

\maketitle

\section{Introduction}

Tin-based binary compounds, Sn$_x$Se$_y$ and Sn$_x$S$_y$, have emerged as promising candidates for electronic, optoelectronic, photovoltaic, and thermoelectric  applications \cite{Ke2013, Vidal2012, Sinsermsuksakul2014, Zhao2014, Ding2015, Yu2016, Zhao2016, Dewandre2016} as well as platforms for exploring exotic states of matter \cite{Chen2017, Yu2017, Marini2019, Zhou2020, Wu2019}. Similar to other layered metal chalcogenide materials~\cite{Qi2016, Pan2015, Leroux2015, Sipos2008, Paudyal2020, Ying2018, Calandra2011, Ali2014}, it has been found that under increased pressure and/or temperature these compounds undergo substantial changes in their structural and  electronic properties, special interest being paid to those materials that can host charge density wave (CDW), superconducting, or topologically non-trivial phases~\cite{Chen2017, Yu2017, Marini2019, Zhou2020, Wu2019, Nguyen-Cong2018, Gonzalez2018, Timofeev1997, Neto2001, Kang2015, Valla2004, Zhou2018, Ying2018}.

In the dichalcogenide systems, pressure-induced structural phase transitions have been reported for both SnSe$_2$ and  SnS$_2$. For example, two theoretical studies have predicted that SnSe$_2$ becomes thermodynamically unstable above 18-20~GPa and then decomposes into Sn$_3$Se$_4$ (space group $I\bar{4}3d$) and Se~\cite{Yu2017, Nguyen-Cong2018}.  A similar decomposition has been theoretically shown to take place in the sister compound SnS$_2$, which remains thermodynamically stable up to approximately 28~GPa~\cite{Gonzalez2018}. We would like to note in passing that both 3:4 compounds have been recently predicted to be superconducting with $T_{\rm c}$ values of 3.3-4.7~K at 10~GPa~\cite{Yu2017, Marini2019}, and 21.9~K at 30~GPa~\cite{Gonzalez2018}, respectively. Although the formation of Sn$_3$Se$_4$ and Sn$_3$S$_4$ has been reported in experiments~\cite{Yu2017, Albers1961},  the presence of a superconducting state in these materials still awaits experimental confirmation. 

\begin{figure*}[t]
	\centering
    \includegraphics[width=0.95\linewidth]{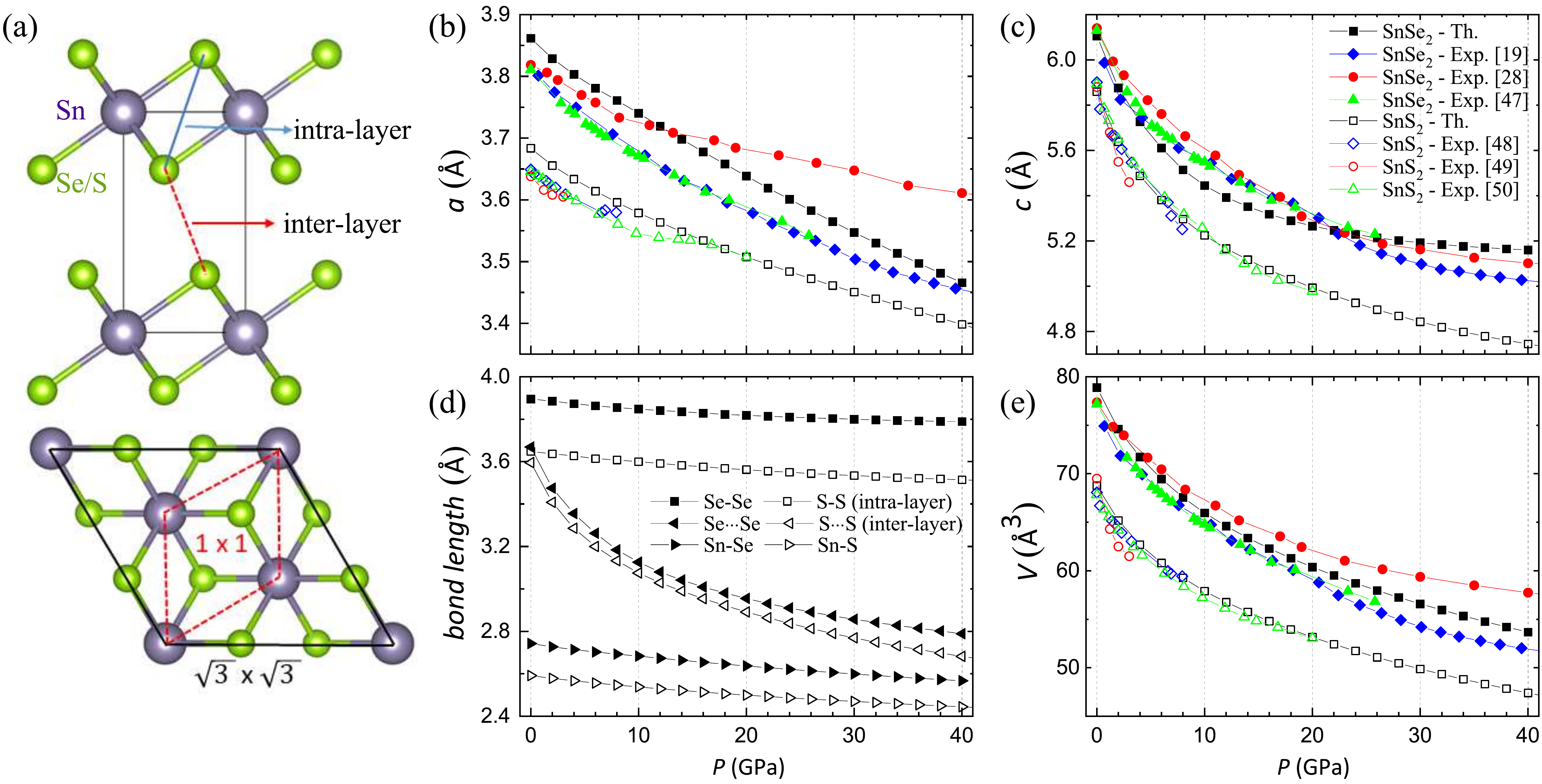}
	\caption{\label{fig:struct_latparam_pressure} (a) Crystal structures showing the intra-layer (Se-Se/S-S) and inter-layer (Se$\cdots$Se/S$\cdots$S) distances (top figure) and $1 \times 1$ and $\sqrt{3} \times \sqrt{3}$ unit cells (bottom figure). Pressure dependence of (b), (c) the lattice parameters $a$ and $c$, (d) the average bond lengths, and (e) the volume per one formula unit for SnSe$_2$ and SnS$_2$. Theoretical results are shown as black symbols and are compared with available experimental data~\cite{Zhou2018, Ying2018, Borges2018, Knorr2001, Hazen1978, Filso2016}. Data for SnSe$_2$ and SnS$_2$ are presented with solid and open symbols.}
\end{figure*}

While SnSe$_2$ is no longer on the convex hull tie-line at higher pressures and also becomes dynamically unstable, we have shown in a previous study~\cite{Ying2018} that it can be stabilized in a $\sq3$ supercell through a periodic lattice distortion (PLD) and that it can, in fact, be realized in experiments. In particular, under applying pressure, SnSe$_2$ becomes metallic in the 8-13~GPa range, with a typical metallic behavior above 17 GPa, and transitions to the PLD phase at around 17~GPa. The agreement between theory and experiment was very good, and the experimentally observed phase transition has been successfully ascribed to a combined effect of electron-phonon (e-ph) coupling and Fermi surface (FS) nesting. This is in contrast to transitional metal dichalcogenides (TMDs), where FS nesting is found to only play a minor role in creating the CDW instability~\cite{Calandra2011, Heil2017, Johannes2008, Calandra2009, Weber2011}. Interestingly, in another recent high-pressure experimental study \cite{Zhou2018}, neither the proposed theoretical decomposition~\cite{Yu2017, Nguyen-Cong2018} nor the transformation to the PLD phase~\cite{Ying2018} were detected up to 46~GPa. Instead, electrical resistance measurements in compressed SnSe$_2$  showed an insulator-to-metallic transition above 15.2~GPa and the appearance  of a superconducting state around 18.6~GPa.   

In the present study, we first address the apparent contradiction among high-pressure experimental results in SnSe$_2$ above 20~GPa. Next, we explore whether a $\sq3$ superstructure can also be stabilized in SnS$_2$ via a PLD by providing a detailed comparison of the electronic, vibrational, and e-ph properties of the two compounds. We expand on our previous work~\cite{Ying2018} to improve the quantitative agreement between theory and experiment with respect to the pressure dependence of the $A_{1g}$ phonon mode.  In particular, we investigate the effect of LO-TO splitting and nonadiabatic corrections on the calculated phonon frequencies. Finally, we investigate the origin of the superconducting state recently observed in SnSe$_2$ under nonhydrostatic pressure, and shed light on the superconducting properties of SnS$_2$ and the PLD phases of the two systems at higher pressures.

\section{Methods}
\label{sec:methods}

First-principles calculations were performed within the density functional theory (DFT) using the Quantum \textsc{Espresso} (QE)~\cite{QE} code. We employed optimized norm-conserving Vanderbilt (ONCV) pseudopotentials~\cite{Hamann2013} with the Perdew-Burke-Ernzerhof (PBE) exchange-correlation functional in the generalized gradient approximation~\cite{PBE}, where the Sn $4d^{10} 5s^2 5p^2$, Se $4s^2 4p^4$, and S $3s^2 3p^4$ orbitals were included as valence electrons. To properly treat the long-range dispersive interactions, we used the non-local van der Waals (vdW) density functional optB86b-vdW~\cite{optB86b, vdW}. A plane wave kinetic-energy cutoff value of 60~Ry, a Marzari-Vanderbilt cold smearing~\cite{Marzari1999} value of 0.01~Ry, and a $\Gamma$-centered $24\times 24 \times 16$ Monkhorst-Pack~\cite{Monkhorst1976} \textbf{k}-mesh for the three-atom unit cell and $12 \times 12 \times 16$ \textbf{k}-mesh for the nine-atom $\sq3$ supercell were used for the Brillouin-zone (BZ) integration. The atomic positions and lattice parameters were optimized until the self-consistent energy was converged within $2.7\times10^{-5}$~eV and the maximum Hellmann-Feynman force on each atom was less than 0.005~eV/\AA. The dynamical matrices and the linear variation of the self-consistent potential were calculated within density-functional perturbation theory (DFPT)~\cite{Baroni2001} on the irreducible set of a regular $6 \times 6 \times 4$ \textbf{q}-mesh  for the three-atom unit cell and $3 \times 3 \times 4$ \textbf{q}-mesh for the nine-atom $\sq3$ supercell. 

The EPW code~\cite{Giustino2007, EPW} was used to compute e-ph interactions and related properties. The electronic wavefunctions required for the Wannier-Fourier interpolation~\cite{WANN1, WANN2} were calculated on a uniform $\Gamma$-centered $12 \times 12 \times 4$ \textbf{k}-grid for the three-atom unit cell. Ten maximally localized Wannier functions (one $s$ and three $p$ orbitals for each Sn atom and three $p$ orbitals for each chalcogen atom) were used to describe the electronic structure near the Fermi level ($E_{\rm F}$). A uniform $300 \times 300 \times 200$ \textbf{k}-mesh was used to evaluate  the adiabatic phonon self-energy and static bare susceptibility, while 2 million random \textbf{k} points were used to estimate the phonon spectral function in the nonadiabatic regime. Both sets of calculations were performed along a high-symmetry path in \textbf{q}-space with smearings of 25~meV (electrons) and 0.05~meV (phonons). Uniform $120 \times 120 \times 40$ \textbf{k}-point and  $60 \times 60 \times 20$ \textbf{q}-point grids were used for the superconductivity calculations in the three-atom unit cell (the in-plane meshes were reduced by a factor of three in the nine-atom supercell). The Matsubara frequency cutoff was set to 0.4~eV and the Dirac deltas were replaced by Lorentzians of width 25~meV (electrons) and 0.1~meV (phonons) when solving the isotropic Migdal-Eliashberg equations. 

\section{Results}
\label{sec:results}
\subsection{Crystal structure evolution under pressure}

\begin{figure}[t]
	\centering
	\includegraphics[width=0.95\linewidth]{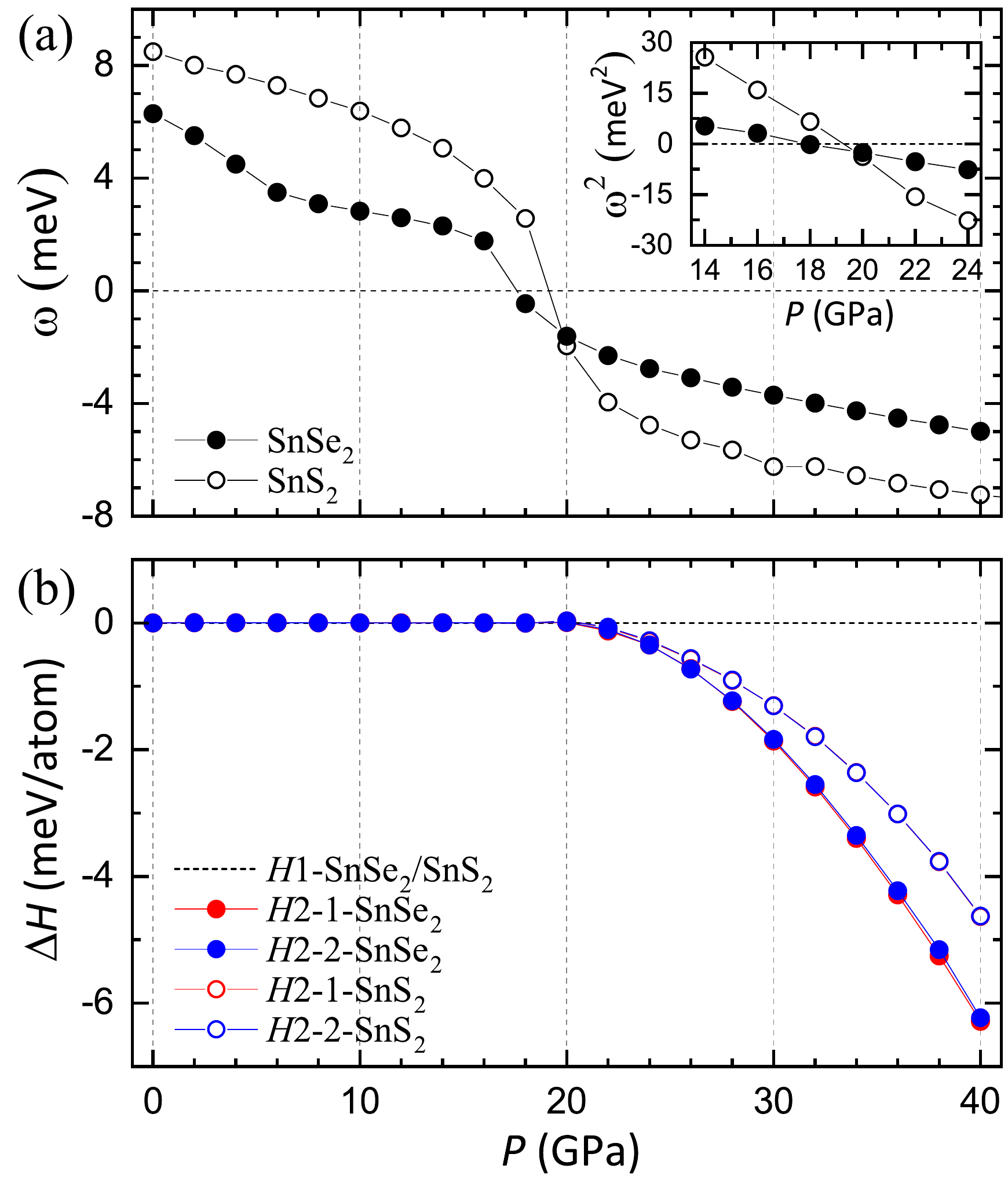}
	\caption{\label{fig:soft_mode_enthalpy} (a) Calculated softening of the lowest-energy (degenerate) phonon modes in the $\sq3$ $H$1 superlattice for SnSe$_2$ and SnS$_2$, where imaginary phonon frequencies are shown as negative, illustrating the pressure-induced destabilization of the $H$1 phase. The inset shows the squared frequency of the modes. (b) Enthalpy difference for the considered $H$2 structures as function of pressure for SnSe$_2$ and SnS$_2$, where the $H$1 phase was chosen as a reference.}
\end{figure}

At ambient conditions, bulk SnSe$_2$ and SnS$_2$ crystallize in the hexagonal, close-packed CdI$_2$-type structure with space group $P\bar{3}m1$ (No.~164)~\cite{Huang2014, Ying2018}. The unit cell contains three atoms, where every Sn atom occupies the center of an octahedron formed by six chalcogen atoms (Se or S) [see Fig.~\ref{fig:struct_latparam_pressure}(a)]. We refer to this high-symmetry structure as the $H$1 phase. Fig.~\ref{fig:struct_latparam_pressure}(b)-(e) presents the optimized structural parameters as a function of pressure, together with available experimental data for comparison~\cite{Zhou2018, Ying2018, Borges2018, Knorr2001, Hazen1978, Filso2016}. 

\begin{figure*}[t]
	\centering
	\includegraphics[width=0.9\linewidth]{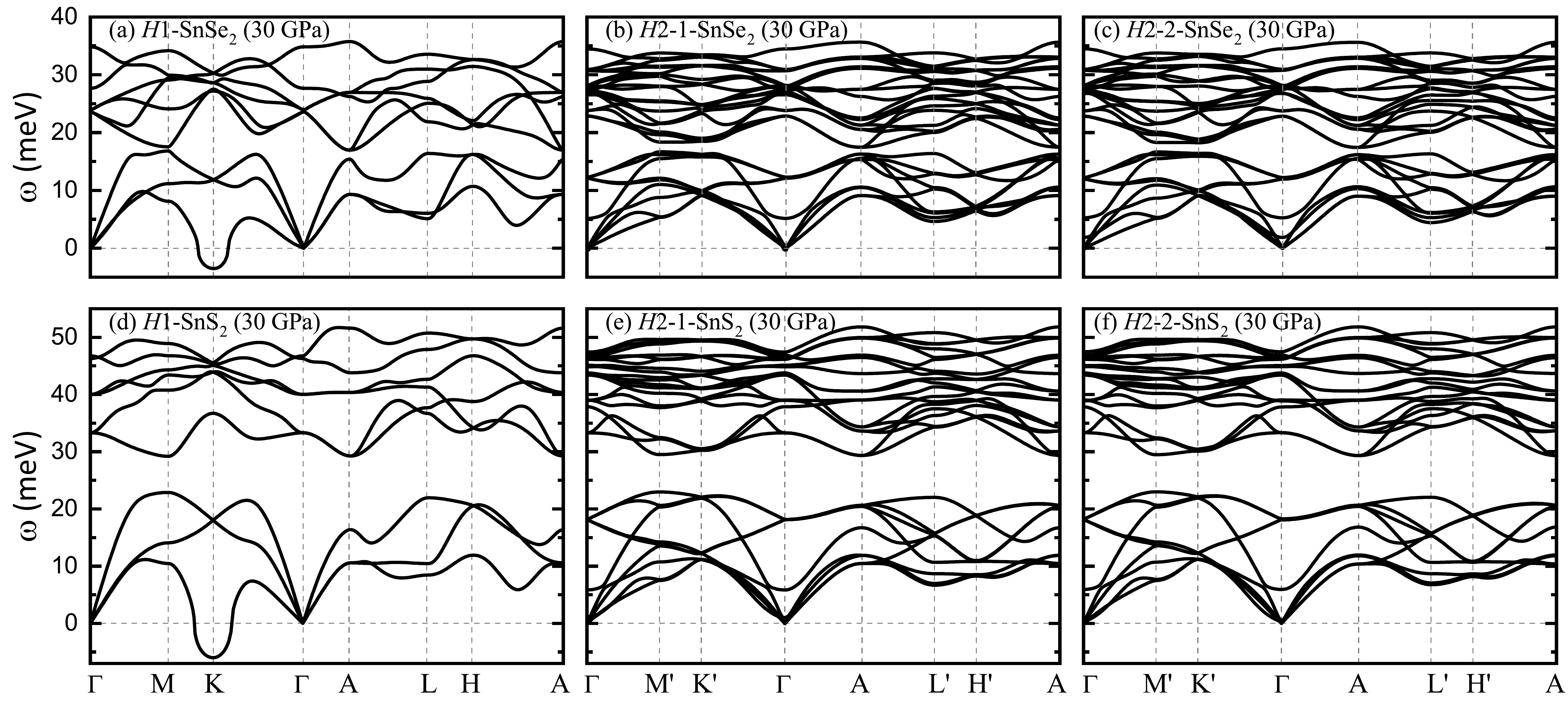}
	\caption{\label{fig:ph-h2} Calculated phonon dispersion for the (a), (d)  $H$1 structure in the $\s1$ unit cell and (b), (c), (e), (f) $H$2 structures in the $\sq3$ supercell of SnSe$_2$ (top row) and SnS$_2$ (bottom row) at 30~GPa. The $H$2-1 and  $H$2-2 structures are derivatives of the undistorted $\sq3$ $H$1 structure created via a PLD.}
\end{figure*}

As pointed out in previous studies \cite{Hazen1978, Knorr2001, Filso2016}, the contraction of the unit cell under compression is highly anisotropic. According to our theoretical results for hydrostatic pressure, the lattice parameter along the $a$ axis decreases almost linearly with increasing pressure, while that along the $c$ axis shrinks quickly before beginning a slower descent. This anisotropic compression can be related to the rapid decrease in the inter-layer Se$\cdots$Se/S$\cdots$S distance (i.e., the distance between the chalcogen atoms in adjacent layers)  versus the intra-layer Se-Se/S-S and Sn-Se/Sn-S  distances due to the weaker vdW  inter-layer interaction relative to the stronger covalent intra-layer bonding. 

Up to 10~GPa, there are no appreciable differences seen in the compressibility rates in the two systems. For instance, the lengths of the $a$ and $c$ axes at 10~GPa are reduced by about 3\% and 11\% with respect to their zero-pressure values in both cases.  As the pressure increases, the compressibility rate along the $a$ axis becomes slightly smaller in SnS$_2$ than in SnSe$_2$, following the expected trend that a more covalent intra-layer bond is less compressible~\cite{Borges2018}~[see Supplemental Fig.~S1~\cite{SM}]. The compressibility rate along the $c$ axis, on the other hand, displays the opposite behavior. This can be empirically understood from the relatively more localized nature of the $3p$ orbitals of S$^{-2}$ compared to the $4p$ orbitals of Se$^{-2}$, which gives rise to a weaker interaction across the vdW gap in SnS$_2$ and thus increased compressibility. 

The present theoretical results for the pressure dependence of the cell parameters are in very good agreement with all experimental data in the low pressure region between 0 and 10~GPa [see Fig.~\ref{fig:struct_latparam_pressure}]. Beyond this point, the pressure transmitting medium used in various experiments can cause considerable differences between the compressed lattice parameters. The largest deviation is observed for the compression of the $a$ axis in SnSe$_2$ beyond 10~GPa  in Ref.~[\onlinecite{Zhou2018}] and is attributed to nonhydrostatic pressure conditions in the experimental setup.  A similar, but smaller effect, was also found in SnS$_2$~\cite{Knorr2001, Filso2016}. 

We next discuss the  thermodynamic and dynamic stability of SnSe$_2$ and SnS$_2$ under compression. With respect to the full phase diagram, our calculations agree well with the findings reported in literature, namely that SnSe$_2$ and SnS$_2$ lie above the convex hull tie-line for pressures above approximately 20 and 40~GPa, respectively [see Supplemental Fig.~S2~\cite{SM}]. The occurrence of a lattice instability is established by calculating the full phonon dispersion relations in the harmonic approximation in the three-atom $H$1 unit cell. As shown in Supplemental Fig.~S3~\cite{SM}, the lowest-energy vibrational mode has an imaginary frequency at the $K$ point of the BZ  at 20~GPa, indicating that the two systems have become dynamically unstable. Based on the above results, it would be expected that the systems will either decompose into more stable products or undergo a crystal transformation into a nearby minima in the configuration space by following the eigenvector of a soft phonon mode. 

The successful synthesis of Sn$_3$Se$_4$ in laser-heated diamond anvil cells~\cite{Yu2017} points towards the first scenario, but no traces of the Sn$_3$Se$_4$ phase were detected in the high-pressure synchrotron x-ray diffraction (XRD) patterns measured at room temperature in two other recent studies~\cite{Ying2018, Zhou2018}. This implies that a relatively large activation barrier needs to be overcome for the system to decompose, requiring not only high pressures but also high temperatures.   

\begin{figure}[t]
	\centering
	\includegraphics[width=0.9\linewidth]{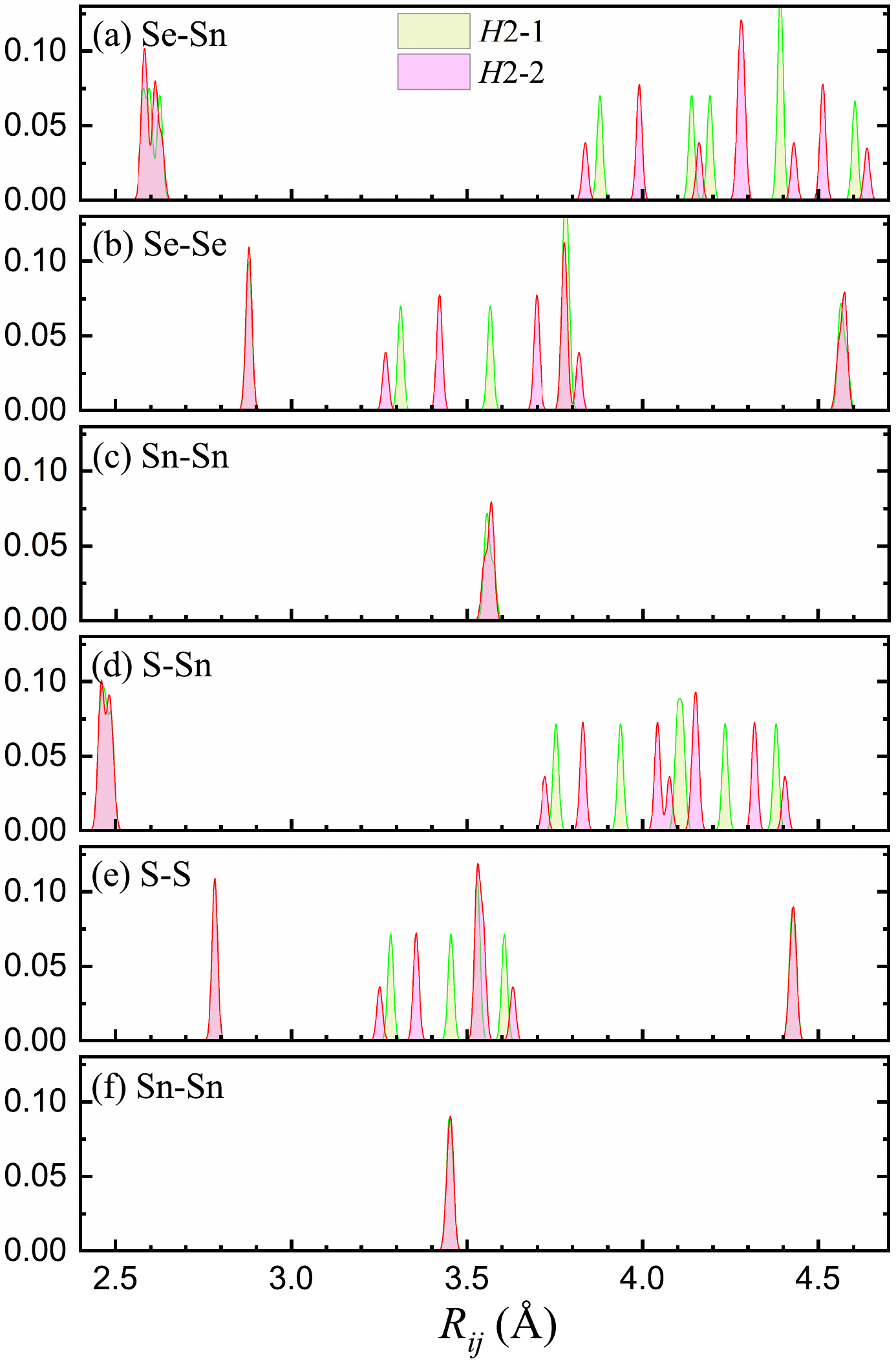}
	\caption{\label{fig:RDF} Calculated radial distribution function of the $H$2 structures for (a)-(c) SnSe$_2$ and (d)-(f) SnS$_2$ at 30 GPa with MAISE~\cite{Hajinazar2020}.}
\end{figure}

\begin{figure*}[t]
	\centering
	\includegraphics[width=0.9\linewidth]{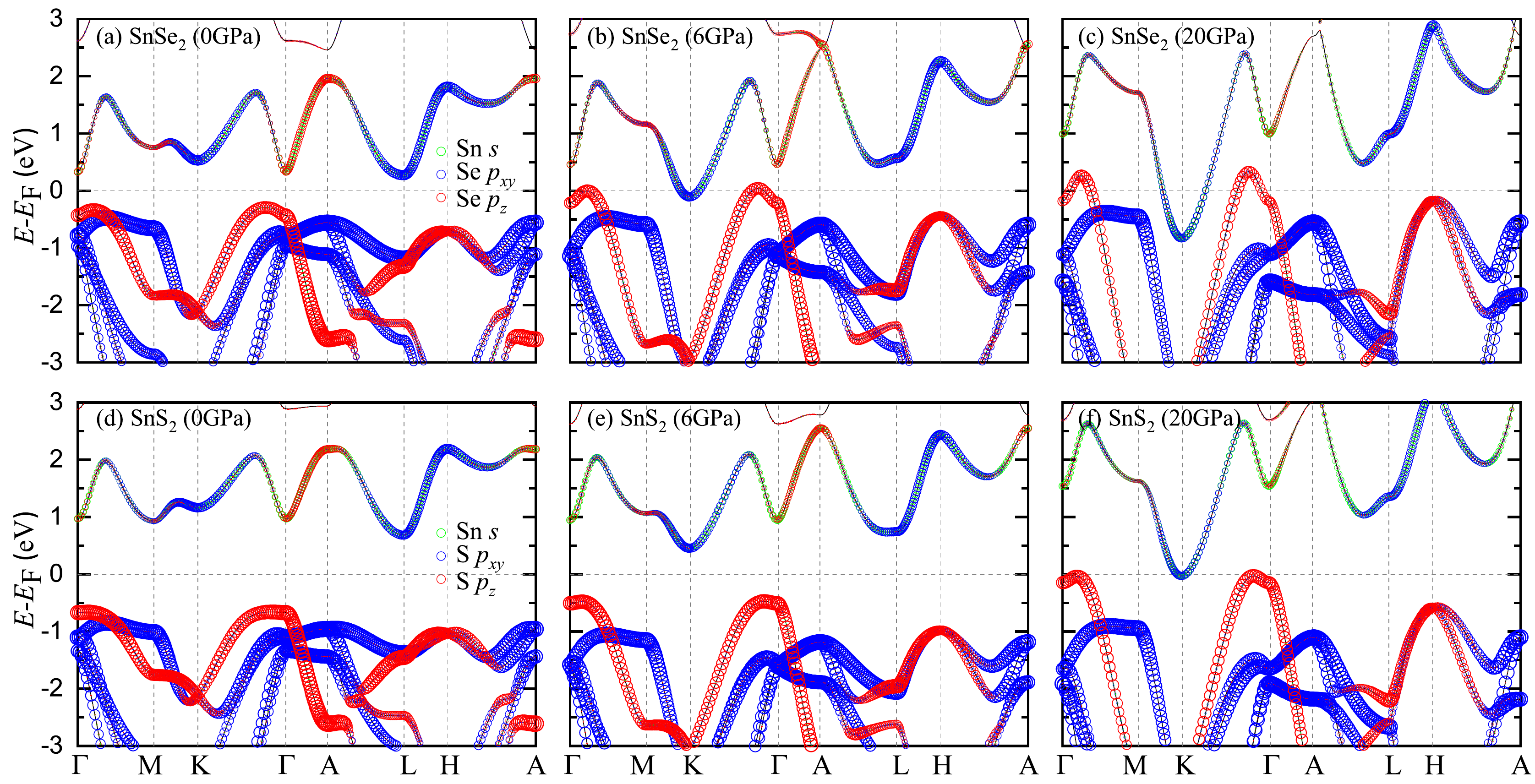}
	\caption{\label{fig:band_structures} Calculated band structure for the $H$1 structure in the $\s1$ unit cell of (a)-(c) SnSe$_2$ and (d)-(f) SnS$_2$ at 0, 6, and 20~GPa. The size of the symbols is proportional to the contribution of each orbital character.}
\end{figure*}

\begin{figure*}[t]
	\centering
	\includegraphics[width=0.9\linewidth]{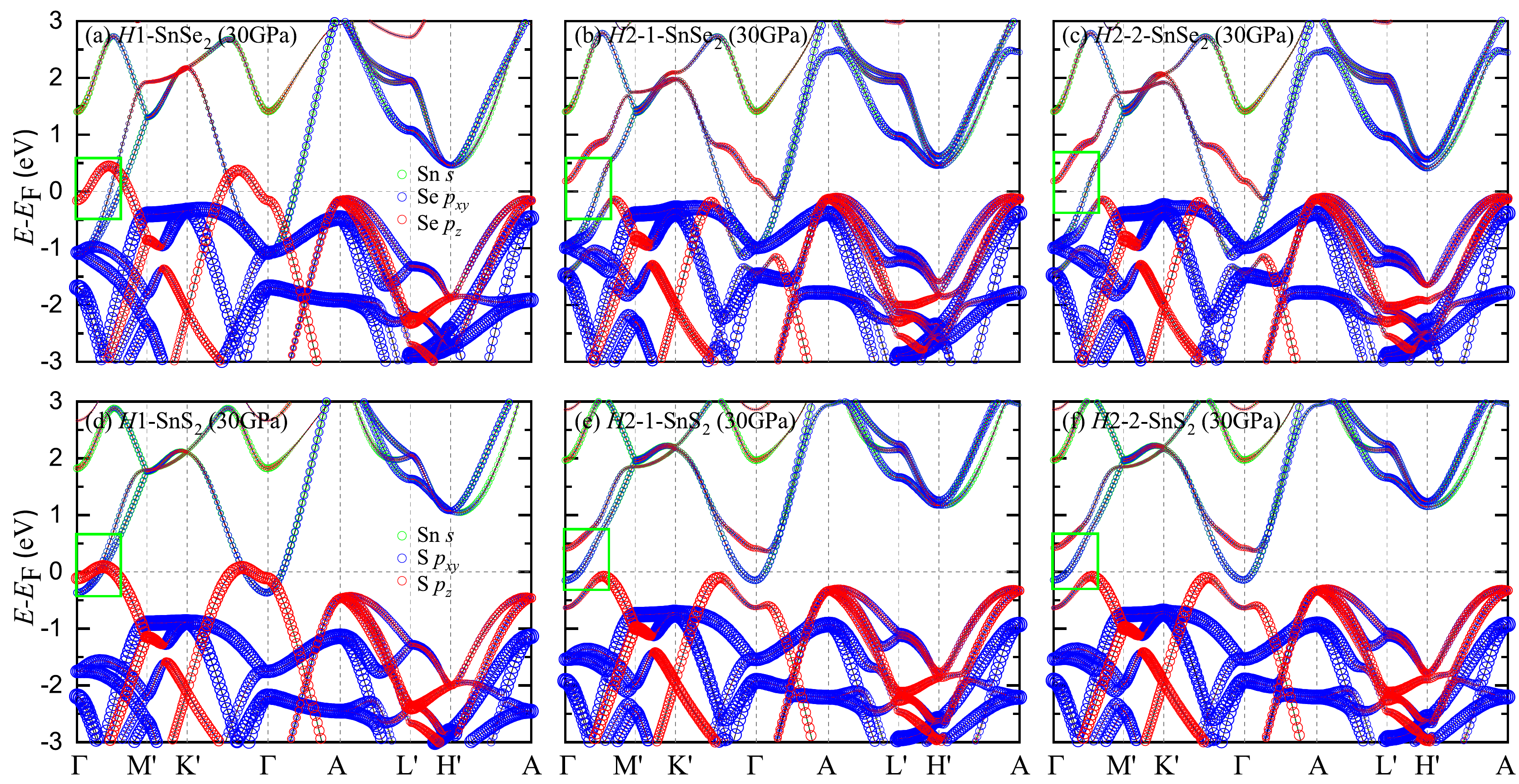}
	\caption{\label{fig:band_structures2} Calculated band structure for the (a), (d)  $H$1 and (b), (c), (e), (f) $H$2 structures in the $\sq3$ supercell of SnSe$_2$ (top row) and SnS$_2$ (bottom row) at 30~GPa. The $H$2-1 and  $H$2-2 structures are derivatives of the undistorted $\sq3$ $H$1 structure created via a PLD. The area indicated by the green box in (a), (d) is to be compared to (b), (e) and (c), (f) where avoided crossing near the Fermi level can be observed in the $H$2 structures.} 	
\end{figure*}

Evidence for the second scenario has also been provided in our combined experimental and theoretical work on SnSe$_2$~\cite{Ying2018}. Namely, the formation of a $\sq3$ superlattice has been revealed through the sudden appearance of several reflections at $\inv3$ in XRD patterns and of two new peaks in the Raman spectra above 17~GPa. In addition, it has been shown that energetically more stable structures with lower symmetry can be constructed as detailed below.  Here we will refer to the derivatives of the undistorted $\sq3$ $H$1 structure created via a PLD as $H$2. The fact that SnSe$_2$ has not been synthesized in the $H$2 structure in the study by Zhou {\it et al.}~\cite{Zhou2018} suggests that the formation of this metastable phase is strongly dependent on the experimental conditions. While the NaCl powder used in our high-pressure XRD  experiments ensured a quasi-hydrostatic pressure environment below 40~GPa~\cite{Ying2018}, the silicone oil used as a pressure transmitting medium by Zhou {\it et al.} produced nonhydrostatic pressures above 13.2~GPa~\cite{Zhou2018}.
A similar situation has been observed for TMDs, where different pressure conditions give rise to variations in the structural and electronic properties.~\cite{Yang2019, Duwal2016}. For instance, in the case of WS$_2$,  the transition from the 2$H_c$ to 2$H_a$ phase occurred under nonhydrostatic, but not under hydrostatic pressure~\cite{Duwal2016}.   

We now focus our attention on understanding whether a $\sq3$ superstructure can also be stabilized in SnS$_2$ via a PLD.  Since the $K$ point of {the $\s1$ unit cell} folds onto the $\Gamma$ point in the $\sq3$ supercell, dynamical stability calculations can be performed by determining the phonon frequencies at the $\Gamma$ point of the nine-atom $H$1 structure. As pressure is increased, the two nearly degenerate lowest-energy phonon modes with $A_{2g}$ and $A_{2u}$ symmetry soften and become imaginary as shown in Fig.~\ref{fig:soft_mode_enthalpy}(a).  The linear evolution of $\omega^2$ with pressure [Fig.~\ref{fig:soft_mode_enthalpy}(a) inset] is a characteristic feature of a soft-mode phase transition~\cite{Fujimoto2005, Kolmogorov2012}, enabling us to estimate the critical transition pressure when $\omega^2$ goes to zero. In this case, we obtain 18~GPa for SnSe$_2$ and 19~GPa for SnS$_2$, respectively. 

To construct the energetically preferred structural derivatives of the original $H$1 phase, we follow the same strategy employed in our previous work on SnSe$_2$~\cite{Ying2018}. We displace the atomic coordinates according to the eigenvectors of $A_{2g}$ ($H$2-1 phase) and $A_{2u}$ ($H$2-2 phase) modes as well as a linear combination of the two eigenvectors ($H$2-3 phase) to better explore the adiabatic potential energy surface. As for SnSe$_2$, the resulting distorted structures for the $\sq3$ supercell are fully relaxed and found to converge to three distinct configurations based on the structure analysis performed with the MAISE package~\cite{Hajinazar2020}. The lattice parameters, space groups, and Wyckoff positions of $H$2 structures at 30~GPa are given in Table.~S1 and the crystal structures are shown in Fig.~S4~\cite{SM}. Since $H$2-3 is an intermediate phase along the pathway that transforms $H$2-1 into $H$2-2, we will only concentrate on the $H$2-1 and $H$2-2 structures in our further discussion. 

As can be appreciated from the pressure dependence plots of the relative enthalpy in Fig.~\ref{fig:soft_mode_enthalpy}(b), the $H2$ phases are energetically more favorable above 18 and 20~GPa for SnSe$_2$ and SnS$_2$, respectively. This is in good agreement with our phonon calculations at 30~GPa, showing no imaginary frequencies and demonstrating the dynamical stability of the predicted $H$2 phases [Fig.~\ref{fig:ph-h2}]. We find the $H2$ derivatives to be virtually degenerate in enthalpy and no distinguishable differences between their phonon spectra. A comparison of the radial distribution functions (RDFs) shown in Fig.~\ref{fig:RDF} along with an estimate of the similarity factor (defined as the RDF dot product between two structures) provide further evidence that the $H$2-1 and $H$2-2 configurations are indeed distinct despite having similar enthalpies. In this case, using a Gaussian spread of 0.008~\cite{Hajinazar2020}, we find a similarity factor of 0.5358 and 0.6102 between the $H$2-1 and $H$2-2 structures for the SnSe$_2$ and SnS$_2$ systems, respectively. Noticeably, the  main difference in the RDFs comes from the second-nearest neighbor, explaining the similar covalent network and, therefore, the very similar vibrational and low-energy electron properties of the $H$2 derivatives. Based on the RDF analysis, we can establish that the distorted configurations are nearby local minima, a situation encountered in other systems~\cite{Shah2013}.

\subsection{Electronic properties}

In this section we systematically analyze the band structure of SnSe$_2$ and SnS$_2$ under hydrostatic and nonhydrostatic pressure. At zero pressure, both systems are found to be semiconducting with an indirect band gap of 0.62~eV in SnSe$_2$ and 1.35~eV in SnS$_2$, respectively. As for bulk TMDs consisting of the same transition metal, the compound with the most electronegative chalcogen has the largest band gap~\cite{Gusakova2017, Zunger1978, DiSalvo1976, Reshak2003}.  While this is the correct trend, we need to keep in mind in our following discussion that the size of the band gaps are underestimated relative to those extracted from experiments~\cite{Domingo1966, Burton2016, Manou1996, Kumagai2016, Kudrynskyi2020} or calculated with the {\it GW} approximation or the HSE06 hybrid functional~\cite{Kumagai2016, Gonzalez2016}. 

Figure~\ref{fig:band_structures} shows the calculated hydrostatic band structures for the $H$1 phase in the three-atom unit cell at 0, 6, and 20~GPa; band structures at other pressures are given in Supplemental Fig.~S5~\cite{SM}. The bottom of the conduction band displays a mixture of Sn $s$ with chalcogen (Se or S) $p_{xy}$ orbitals along the in-plane directions and with chalcogen $p_{z}$ orbitals along the out-of-plane directions. The top of the valence band, on the other hand,  consists almost entirely of chalcogen $p_z$ orbitals. Under compression, the orbital character of the bands remains largely unaffected, but the bandwidths expand as indicated by the increase in the slope of the dispersion curves. Near the Fermi level, the most significant change takes place in the lowest conduction band level along the $\Gamma-K$ segment.  Due to a greater overlap between the chalcogen $p_{xy}$ and Sn $s$ orbitals, the band energy at the $K$ point decreases, eventually dropping below the valence band maximum and closing the band gap. As a result, a pressure-induced semiconductor-to-metal transition is estimated to occur at 6 and 20~GPa in SnSe$_2$ and SnS$_2$, respectively. 

The predicted metallization pressures are consistent with available theoretical results~\cite{Nguyen-Cong2018, Ying2018,Filso2016}, but lower than the experimental values as anticipated  from the underestimation of the band gaps in the PBE approximation. Additional changes in the metallization pressure are also expected under nonhydrostatic pressure conditions. DFT results show that a transition at a higher (lower) pressure is favored under in-plane (out-of-plane) uniaxial compressive strain [see Supplemental Fig.~S6 \cite{SM}]. This trend is in agreement with experimental resistivity measurements in SnSe$_2$ where a semiconductor-to-metal transition was observed between 8-13~GPa under quasi-hydrostatic pressure~\cite{Ying2018} compared to a transition above 15.2~GPa under nonhydrostatic pressure~\cite{Zhou2018}. As can be seen in Fig.~\ref{fig:struct_latparam_pressure}(b), the in-plane compression in the latter study is considerable smaller, thus pushing the metallization point to a higher pressure. 

To get a better understanding of how the internal structural parameters of the layers, the inter-layer distance, and the choice of chalcogen atom affect the electronic transition, we calculate the electronic dispersions of (i) SnSe$_2$ structure at 0~GPa in which the Se atoms are substituted with S and the atomic coordinates are either kept unchanged (labeled as SnSe$_2$-str-S) or allowed to relax (labeled as SnSe$_2$-str-S-relaxed), and (ii) SnS$_2$ structure at 0~GPa in which the S atoms are substituted with Se and the atomic coordinates are either kept unchanged (labeled as SnS$_2$-str-Se) or allowed to relax (labeled as SnS$_2$-str-Se-relaxed). 

As shown in Supplemental Fig.~S7 \cite{SM}, the effect of replacing Se with S is an up-shift of the lowest conduction band states and, consequently, a slight increase in the band gap from 0.62 to 0.75~eV. Allowing the atoms to relax gives an additional up-shift of 0.15~eV since the intra-layer distance is reduced to almost the ideal value in SnS$_2$. However this up-shift is still not large enough to reach the band gap value of SnS$_2$. If, on the other hand, we use the SnS$_2$ structure with the Se pseudopotential, as in calculation (ii), we see an opposite trend since now the lowest conduction band states move down in energy. The band gap reduces from 1.35 to 0.90~eV without atomic relaxation and further to 0.22~eV once the atoms are allowed to relax. The reduction by a factor of almost three of the band gap relative to the value in ideal SnSe$_2$ can be ascribed to the presence of slightly more charge in the vdW gap as the inter-layer distance is compressed by almost 10\%. Supplemental Figs.~S8 and S9~\cite{SM} show total charge and charge redistribution plots for the configurations described above. Overall, the metallization process is driven by the cooperative effect of both chemical and structural factors. 

We further study the evolution of the charge density with pressure. Similar to TMDs, we find that under compression more electronic charge moves away from the Sn atoms and accumulates along the intra-layer bonds formed by Sn with the chalcogen atoms and in the inter-layer region between the chalcogen atoms~\cite{Nayak2014, Guo2013, Rajaji2018} [Supplemental Figs.~S10~\cite{SM}]. This view is also supported by the increase in the in-plane average charge at the middle of the vdW gap under applied pressure [see Supplemental Figs.~S11 \cite{SM} showing the in-plane average charge as a function of the perpendicular direction with respect to the layer]. This behavior is in contrast to intercalated carbon and boron layered compounds where, under applied pressure, the inter-layer charge from the intercalant atoms is forced out to the covalent sheets~\cite{Zhang2006, Kim2006, Calandra2007-2}.

Finally, in Fig.~\ref{fig:band_structures2} we compare the electronic structure of $H$1 and $H$2 phases in the $\sq3$ supercell for the two systems.  As shown in our previous study on SnSe$_2$~\cite{Ying2018}, the main difference lies in the lifting of electronic degeneracy near the Fermi level along the $\Gamma-M$ and $K-\Gamma-A$ directions in the $H$2 structures. Compared to the parent $H$1 phase, there is an out-of-plane displacement of the Sn atoms resulting in a slight buckling of the Sn layers and a shift of the chalcogen atoms with respect to the center of the octahedron [see Supplemental Fig.~S4~\cite{SM}]. This modulation of atomic positions leads to the avoidance of crossings between the bands with mixed Sn $s$ and chalcogen $p_{xy}$ orbitals and the band with chalcogen $p_z$ character. Similarly to the phonon spectra in Fig.~\ref{fig:ph-h2}, there are no noticeable differences between the band structure plots of the $H2$ derivatives.

\subsection{Vibrational properties}

In low-dimensional or layered systems, the appearance of a superlattice is often the signature of a CDW transition, as for example in TMDs~\cite{Johannes2008, Calandra2009}. In these materials, it has been established that the wave vector dependence of the e-ph coupling is crucial in understanding the CDW formation, while the FS nesting has been found to only play a minor role \cite{Calandra2011, Heil2017, Johannes2008, Calandra2009, Weber2011}. As presented in our previous study on SnSe$_2$~\cite{Ying2018}, the breaking of electronic degeneracies by a phonon-modulated lattice distortion, the reduction of the density of states (DOS) at the Fermi level, and the softening of a low-energy phonon are key signatures of a momentum-dependent e-ph coupling CDW instability. In this section, we want to investigate if the same reasoning is true for SnS$_2$ as well.

To address the microscopic mechanism responsible for the phase transition and the origin of phonon softening at $K$, we calculate the adiabatic phonon self-energy ($\pqva$) for the lowest-energy phonon mode and the static bare susceptibility in the constant matrix approximation ($\chi^{0, \rm {CMA}}_{\bq}$) using the following equations \cite{Giustino2017, EPW, Zhang2005}

\begin{eqnarray}
\label{eq:PiA}
\pqva &=& 2\summn \intbz\left[\frac{\fnk-\fmkq}{\emkq-\enk}\right] \times \nonumber \\& &
 \times \ \gmnvkqb \times \gmnvkqs   \\
\label{eq:chi0}
\chi^{0, \rm {CMA}}_{\bq} &=& 2\summn \intbz  \left[\frac{\fnk-\fmkq}{\emkq-\enk}\right]
\end{eqnarray}

\begin{figure}[t]
	\centering
	\includegraphics[width=\linewidth]{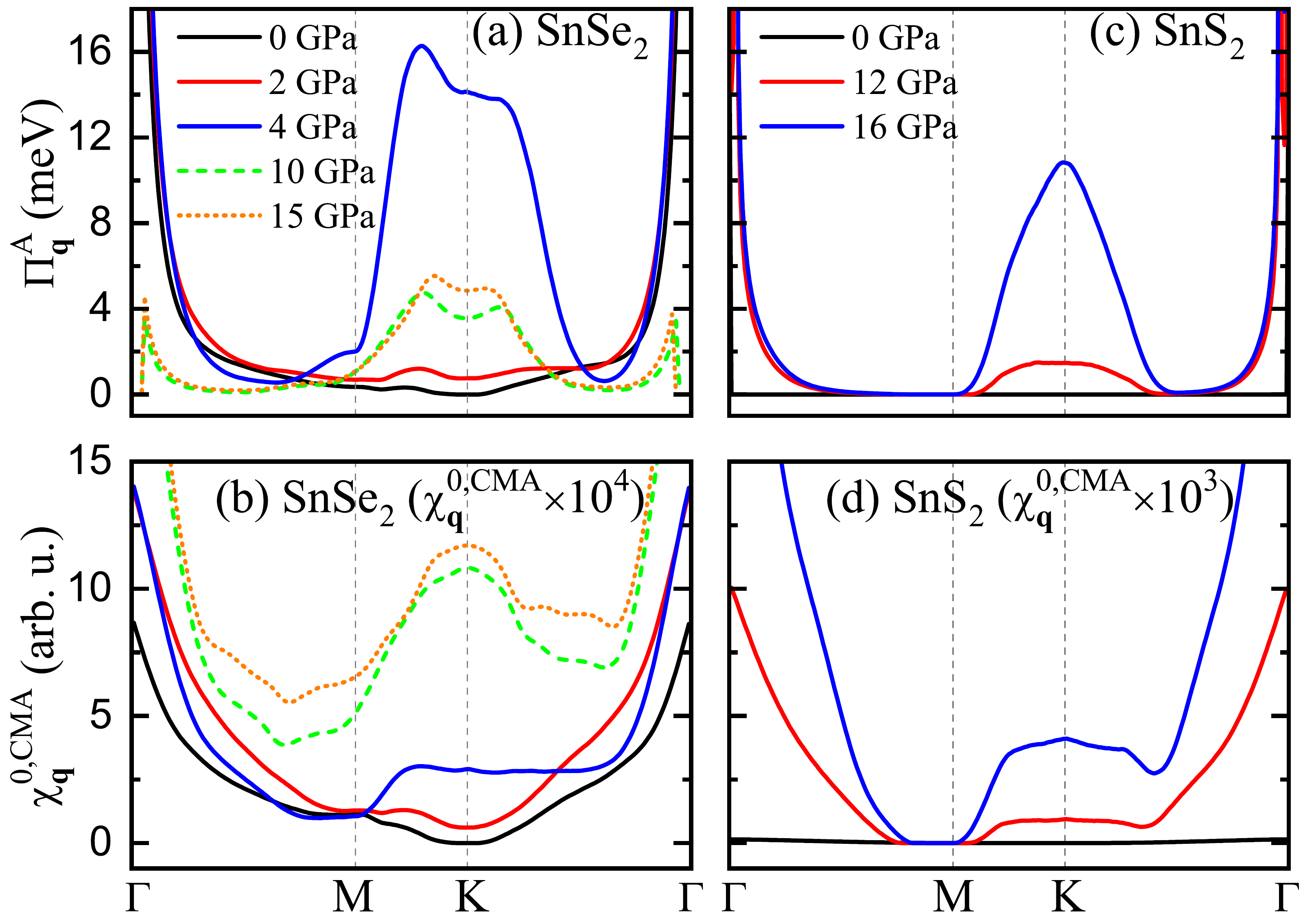}
	\caption{\label{fig:Pi0_chi0} The adiabatic phonon self-energy for the soft phonon mode  and the static bare susceptibility in the constant matrix approximation for the $H$1 structure in the $\s1$ unit cell of (a)-(b) SnSe$_2$ and (c)-(d) SnS$_2$ as a function of pressure along a high-symmetry $\mathbf{q}$-path. The dashed and dotted lines for SnSe$_2$ indicate pressures at which the material is in the metallic phase~\cite{note2}. The $\chi^{0, \rm {CMA}}_{\bq}$ values in (b) and (d) should be multiplied by a factor of 10$^4$ and 10$^3$, respectively.}
\end{figure}

\begin{figure}[t]
	\centering
	\includegraphics[width=0.95\linewidth]{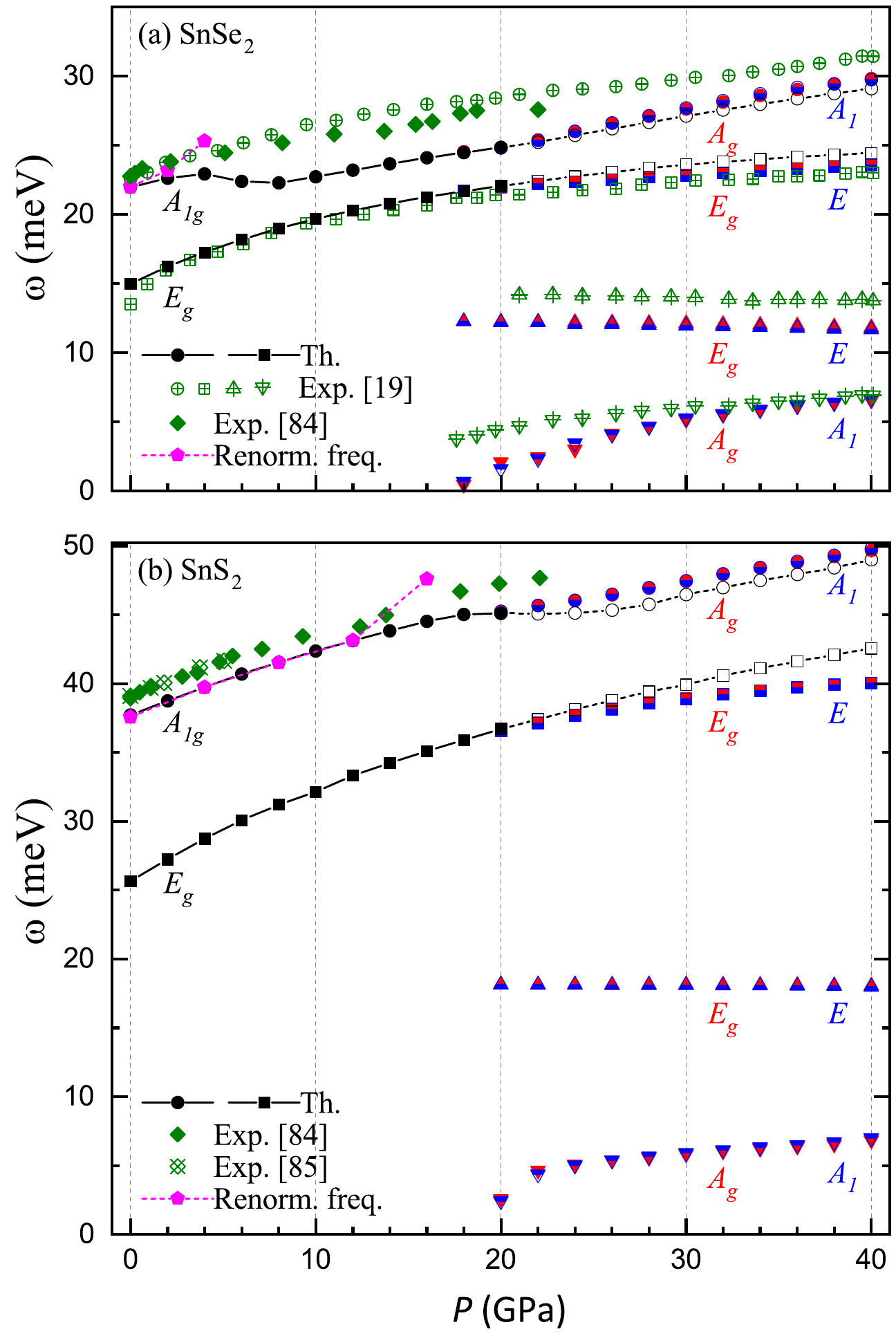}
	\caption{\label{fig:raman} Calculated frequency dependence of the Raman-active modes at $\Gamma$ as a function of pressure for (a) SnSe$_2$ and (b) SnS$_2$.  The theoretical results are shown as black symbols ($H$1), red symbols ($H$2-1), and blue symbols ($H$2-2) while the experimental results~\cite{Ying2018, Bhatt2015, Utyuzh2010} are shown as olive symbols. For $H$1, the data before and after the phase transition are shown as filled and open symbols. The nonadibatically corrected frequency points are shown as filled magenta pentagons.}
\end{figure}

Here, 2 is the spin degeneracy factor, $\enk$  and $\fnk$ represent single-particle energies and Fermi-Dirac occupation factors, respectively, and $\Omega_{\rm {BZ}}$ is the BZ volume. The screened e-ph matrix elements $\gmnvkq$ were obtained as $\gmnvkq = \langle \psimkq|\pv|\psink \rangle/\sqrt{\hbar/2\omega_{\bq\nu}}$, where $\psink$ represents the Kohn-Sham single-particle eigenstates, $\pv$ is the derivative of the self-consistent potential, and $\w_{\bq\nu}$ is the phonon frequency associated with phonon branch $\nu$ and wave vector $\bq$. The bare matrix element $\gmnvkqb$ is calculated by multiplying the screened e-ph matrix elements $\gmnvkq$ with the electronic dielectric function at $\w$=0 [$\ve(\bq, \w= 0)$]. Thus, the matrix elements $\gmnvkqb \times \gmnvkqs$ in Eqn.~(\ref{eq:PiA}) can be replaced by $|\gmnvkq|^2\times~\ve(\bq, \w= 0)$~\cite{note3}.

\begin{figure*}[t]
	\centering
	\includegraphics[width=0.9\linewidth]{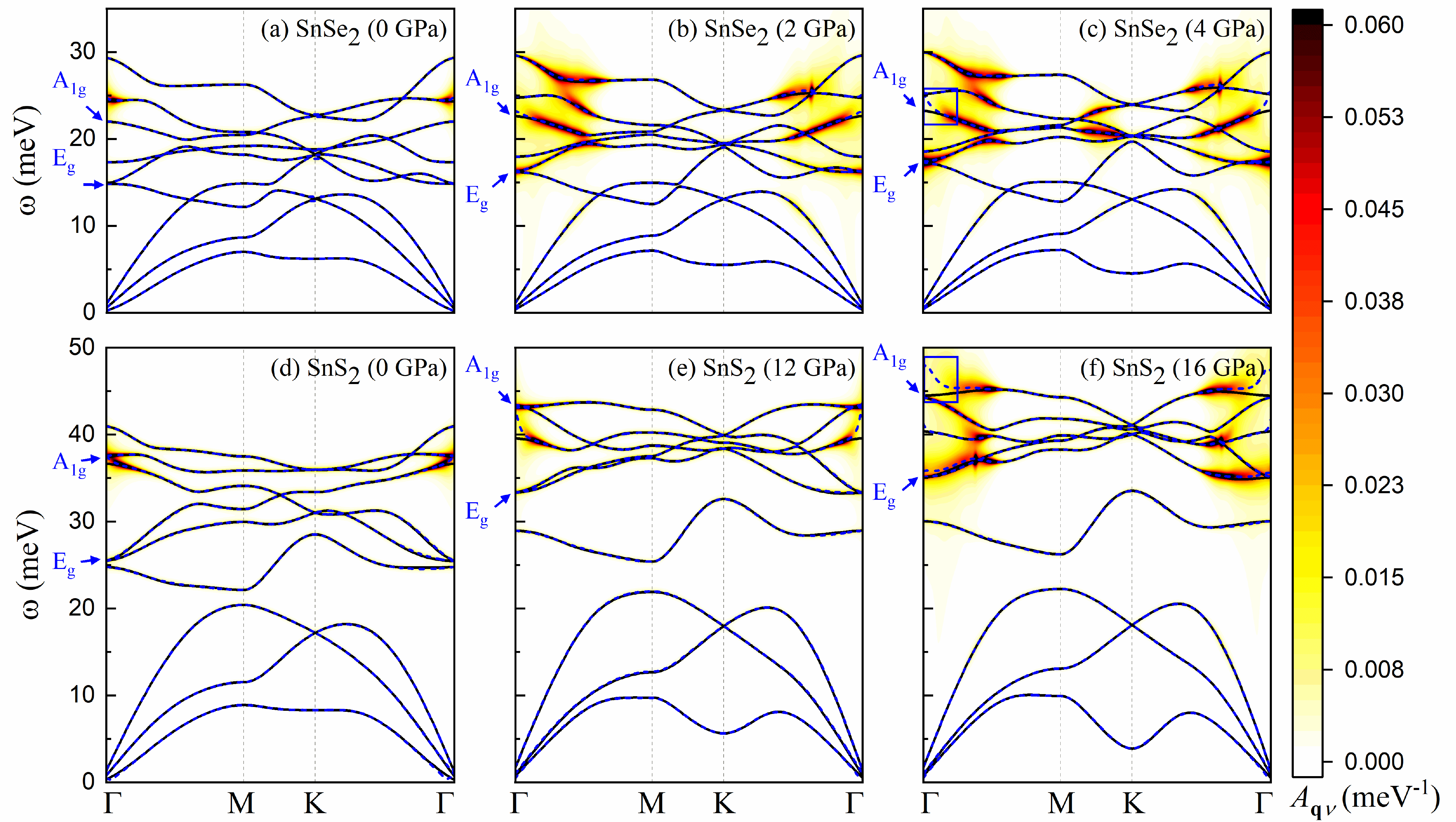}
	\caption{\label{fig:spectral_function} Calculated nonadiabatic phonon spectral function and  nonadiabatic renormalized phonon for the $H$1 structure in the $\s1$ unit cell of (a)-(c) SnSe$_2$ and (d)-(f) SnS$_2$ at various pressures. The solid black and dashed blue lines represent the DFPT phonon and renormalized phonon dispersions, respectively. The color map corresponds to the nonadiabatic phonon spectral function.}
\end{figure*}

Both $\chi^{0, \rm {CMA}}_{\bq}$ and $\pqva$ are properties of the FS bands and those electronic states close to them, yet $\chi^{0, \rm {CMA}}_{\bq}$ is purely electronic, while $\pqva$ includes the fully anisotropic e-ph interaction. As can be appreciated in Fig.~\ref{fig:Pi0_chi0}, we observe an increasing response of $\pqva$ and $\chi^{0, \rm {CMA}}_{\bq}$ at the $K$ point in both systems with increasing pressure, meaning that both FS nesting as well as the e-ph coupling strength increase at the BZ regions close to $K$ as the critical pressure for the phase transition is approached. This leads us to the conclusion that both FS nesting and strong e-ph interactions are providing important contributions in creating the PLD instability in SnS$_2$, in close analogy to SnSe$_2$~\cite{Ying2018, note1, note2}.

Having identified the PLD instability in SnS$_2$ to be of the same origins as in SnSe$_2$, we carry on to compare the materials' Raman-active modes' frequencies as a function of pressure, as shown in Fig.~\ref{fig:raman}. The high energy $E_g$ mode increases monotonically as a function of pressure in both SnSe$_2$ and SnS$_2$. The high energy $A_{1g}$ mode, on the other hand, shows a noticeable change of slope at approximately 6 and 20~GPa in the case of SnSe$_2$ and SnS$_2$, respectively. This can be related to the semiconductor-to-metal transition in both systems as also supported by Supplemental Fig.~S12~\cite{SM}. After the phase transition, due to the in-plane tripling of the unit cell, two additional Raman-active modes appear at 17~GPa for SnSe$_2$ and at 20~GPa for SnS$_2$, in good agreement with available experimental data~\cite{Ying2018}. 

The pressure dependence of the high-energy $E_g$ mode for SnSe$_2$ is in very good agreement with the experimental results, however, the high-energy $A_{1g}$ mode, while reproducing nicely the qualitative trend, underestimates the phonon energies by $\sim 0.5$~meV at 0~GPa and $\sim 3$~meV at 6 GPa (the metallization pressure) compared to experiments~\cite{Ying2018}. Recent studies on other insulators and semiconductors show that the underestimation of phonon energies with respect to experiments can be due to the neglection of nonadiabatic effects in the theoretical approach~\cite{Pisana2007, Caudal2007, Calandra2007, Saitta2008, Calandra2010}. In order to assess the effect of nonadiabatic corrections on the phonon dispersions in our systems, we employed a field-theoretic framework~\cite{Giustino2017}, where the phonon self-energy $\pqv$ can be partitioned into adiabatic and nonadiabatic contributions, i.e., $\Pi(\omega) = \Pi^\text{A} + \Pi^\text{NA}(\omega)$. Thus, the nonadiabatic corrections to the self-energy as a function of the phonon mode $\nu$ and wave vector $\mathbf{q}$ are given by:

\begin{eqnarray}
\label{1}
\pqvna  &=& 2\summn \intbz \times \gmnvkqb \times \gmnvkqs \nonumber \\&\times& \left[\frac{\fmkq-\fnk}{\emkq-\enk-\hbar(\w+i\eta)} -\frac{\fmkq-\fnk}{\emkq-\enk}\right]  \nonumber \\
\end{eqnarray}
with $\eta$ being a positive infinitesimal number. The nonadiabatic phonon spectral function~\cite{Allen1972, Caruso2017} can then be calculated using

\begin{equation}
\label{eq:A_NA}
A^\text{NA}_{\bq\nu}(\w)= \frac{1}{\pi} {\rm {Im}} \left[\frac{2\hbar\w_{\bq\nu}}{{(\hbar\w)}^2 - (\hbar\w_{\bq\nu})^2- 2\hbar\w_{\bq\nu}\pqvnaw }\right]
\end{equation}

Equation~(\ref{eq:A_NA}) reveals that a strong response of the system is expected at the nonadiabatic phonon frequency, $\Omega_{\bq\nu}$, when the denominator is very small or vanishes.
The renormalized nonadiabatic phonon energy, that is, the phonon energy modified by the phonon self-energy of Eqn.~(\ref{1}) is thus given by 

\begin{eqnarray}
\label{eq:Omega}
\Omega^2_{\bq\nu} \simeq \omega^2_{\bq\nu} + 2 \omega_{\bq\nu}{\rm {Re}} \Pi_{\bq \nu}^{\rm {NA}}(\Omega_{\bq\nu})
\end{eqnarray}

The nonadiabatic spectral functions and the nonadiabatic phonon dispersions  obtained via Eqns.~(\ref{eq:A_NA}) and (\ref{eq:Omega}) at various pressures along the $\Gamma-M-K-\Gamma$ direction for SnSe$_2$ and SnS$_2$ are reported in Fig.~\ref{fig:spectral_function}. As can be appreciated from Fig.~\ref{fig:spectral_function}, the nonadiabatic effects are very small in both systems at ambient pressure. As the pressure increases (and the band gap decreases), the renormalization of the $A_{1g}$ mode energy becomes significant around 4~GPa in SnSe$_2$ and around 16~GPa in SnS$_2$.  In particular, the $A_{1g}$ mode in SnSe$_2$ hardens by $\sim 2$~meV at 4~GPa [Fig.~\ref{fig:spectral_function}(c)], improving the quantitative agreement of our calculations with the experiments~\cite{Ying2018} [see also Fig.~\ref{fig:raman}]. Similarly, the $A_{1g}$ mode in SnS$_2$ hardens by $\sim 3$~meV at 16~GPa as shown in Fig.~\ref{fig:spectral_function}(f). For the $E_g$ mode, where the agreement between experiments and theory has already been very good, the nonadiabatic corrections are found to be very small.

As SnSe$_2$ and SnS$_2$ are polar semiconductors at ambient and low pressures, we also want to address the topic of LO-TO splitting. At ambient pressure, we observe a LO-TO splitting of 7~meV for SnSe$_2$ and 11.8~meV for SnS$_2$ [see Supplemental Fig.~S13~\cite{SM}]. With increasing pressure, the band gap decreases and the value for the dielectric function increases. In turn, the magnitude of the LO-TO splitting decreases, in accordance with experimental measurements in other polar semiconductors \cite{Wagner2000, Goni2001}. Interestingly, however, at the critical pressures above which our compounds become metallic, our calculations still yield finite values for the splitting (around 6.4~meV for SnSe$_2$ and 10~meV for SnS$_2$). A possible explanation for the residuary LO-TO splitting is that in the pressure regions, where the materials pass through the semiconductor-metal phase transition, the charge carrier concentration is so low that the long-range Coulomb interactions are not fully screened, thus allowing for finite LO-TO splitting. In the fully metallic state at higher pressures, the long-range Coulomb interactions are screened completely leading to the degeneracy of LO and TO modes. Although highly interesting, resolving the intermediate bad metal phase would go beyond the scope of this work and is therefore the focus of future investigations.

\subsection{Superconducting properties}

\begin{figure*}[t]
	\centering
	\includegraphics[width=0.95\linewidth]{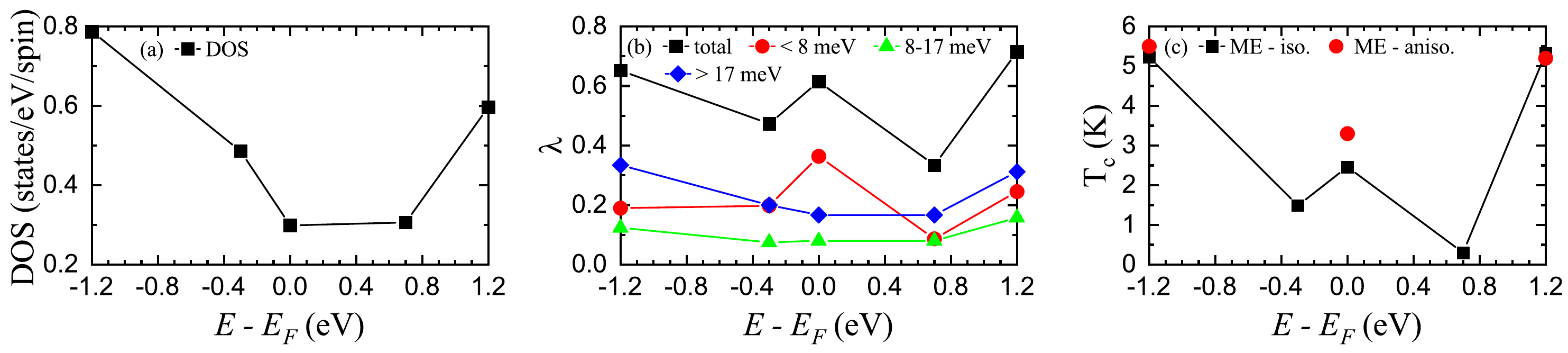}
	\caption{\label{fig:supercond} Variations in (a) DOS at $E_{\rm F}$, (b) $\lambda$, and (c) $T_{\rm c}$ as a function of a rigid shift of the Fermi level  for the $H$1 structure in the $\s1$ unit cell of SnSe$_2$ at the experimental unit cell parameters at 23~GPa.  In (b), squares represent the total $\lambda$, while circles, triangles, and rhombuses represent the contribution of the low-, medium- and high-energy modes. In (c), circles and squares represent the $T_{\rm c}$ obtained from the numerical solutions of the anisotropic and isotropic ME equations.}
\end{figure*}

Experimental evidence of pressure-induce superconductivity in SnSe$_2$ has been recently provided based on electrical transport and synchrotron XRD measurements~\cite{Zhou2018}. The superconducting state is observed to emerge at 18.6~GPa and to reach a maximum $T_{\rm c}$ of about 6.1~K, which remains nearly constant in a large pressure range between 30.1 and 50.3~GPa. Earlier studies have demonstrated superconductivity in bulk and thin films SnSe$_2$ through intercalation and gating~\cite{Ohara1992,Li2017,Song2019,Wu2019} as well as interface superconductivity in SnSe$_2$/ion-liquid and SnSe$_2$/graphene~\cite{Zeng2018, Zhang2018}.

In this section we investigate the origin of the observed superconducting transition in SnSe$_2$ using the Migdal-Eliashberg (ME) formalism implemented in the EPW code~\cite{EPW, Margine2013}.  To facilitate the comparison with the experimental results of Zhou {\it et al.}~\cite{Zhou2018},  the superconducting properties are calculated at the experimental unit cell parameters for nonhydrostatic pressure points at 23 and 30~GPa  while allowing the atomic positions to relax. The resulting electronic structure and phonon dispersion are shown in Supplemental Figs.~S14-S16~\cite{SM}. A comparison with the plots under hydrostatic conditions at 20 and 30~GPa [Figs.~\ref{fig:band_structures}, \ref{fig:band_structures2}, and S3] demonstrate that the electronic structure is highly insensitive to the crystal structure parameters while the phonon frequencies are slightly affected. For instance, according to our phonon calculations within the harmonic approximation a lattice instability starts developing at 20 and 30~GPa under hydrostatic and nonhydrostatic conditions, respectively.  Since the pressure-dependence of the lowest acoustic mode at the $K$ point is responsible for the dynamical instability in this compound, the reduced in-plane compressibility rate in the latter case leads to a decrease in the phonon softening rate and consequently an increase in the stability region up to 30~GPa. While this trend is in the right direction, the  transition pressure is still underestimated compared to the nonhydrostatic experiments in Ref.~[\onlinecite{Zhou2018}] that showed no structural phase transition or decomposition up to 46 GPa. The situation is reminiscent of the one encountered in NbSe$_2$ and NbS$_2$, where the instability found at the harmonic level has been shown to be weakened or even removed when quantum anharmonic effects are taken into account, and therefore suppress the formation of the CDW phase~\cite{Leroux2015,Heil2017}. A similar scenario occurring in SnSe$_2$ would extend the stabilization region to higher pressures as found in experiment. To check this point, we re-evaluated the phonons at the $K$ point with a larger smearing value of 0.03~Ry and indeed found that the imaginary phonon mode is removed, thus preserving the dynamic stability [see dash line in Supplemental Fig.~S16(a)~\cite{SM}].

In order to understand the origin of the superconducting state, we evaluate the Eliashberg spectral function $\alpha^2F(\omega)$ and the corresponding e-ph coupling strength $\lambda$. A comparative analysis of  the $\alpha^2F(\omega)$ and the atomically resolved phonon DOS (PHDOS) at the experimental lattice cell parameters at 23~GPa [Supplemental Fig.~S15~\cite{SM}] reveals that the low-frequency phonons below 8~meV associated with both Sn and Se vibrations contribute 60\%  of the total $\lambda=0.62$, while the phonons in the high-frequency region between 17 and 32~meV dominated by Se atoms vibrations give approximately 30\% of $\lambda$ [Fig.~\ref{fig:supercond}(b)]. Using a typical value of $\mu^*=0.1$, we predict a $T_{\rm c}$ of 2.5 and 3.3~K from the numerical solutions of the isotropic and anisotropic ME gap equations~\cite{Margine2013, EPW}, slightly below the 4.0~K onset temperature ($T\rm{^{onset}_c}$) of electrical resistance drop at 22~GPa reported in Ref.~[\onlinecite{Zhou2018}]. The results are almost unchanged when the calculations are performed at the experimental lattice cell parameters at 30~GPa  [Supplemental Fig.~S16~\cite{SM}]. For the $K$ point phonons computed at the larger smearing value of 0.03~Ry, we get $\lambda=0.59$ and an isotropic $T_{\rm c}=2.3$~K with $\mu^*=0.1$. 

\begin{figure*}[t]
	\centering
	\includegraphics[width=\linewidth]{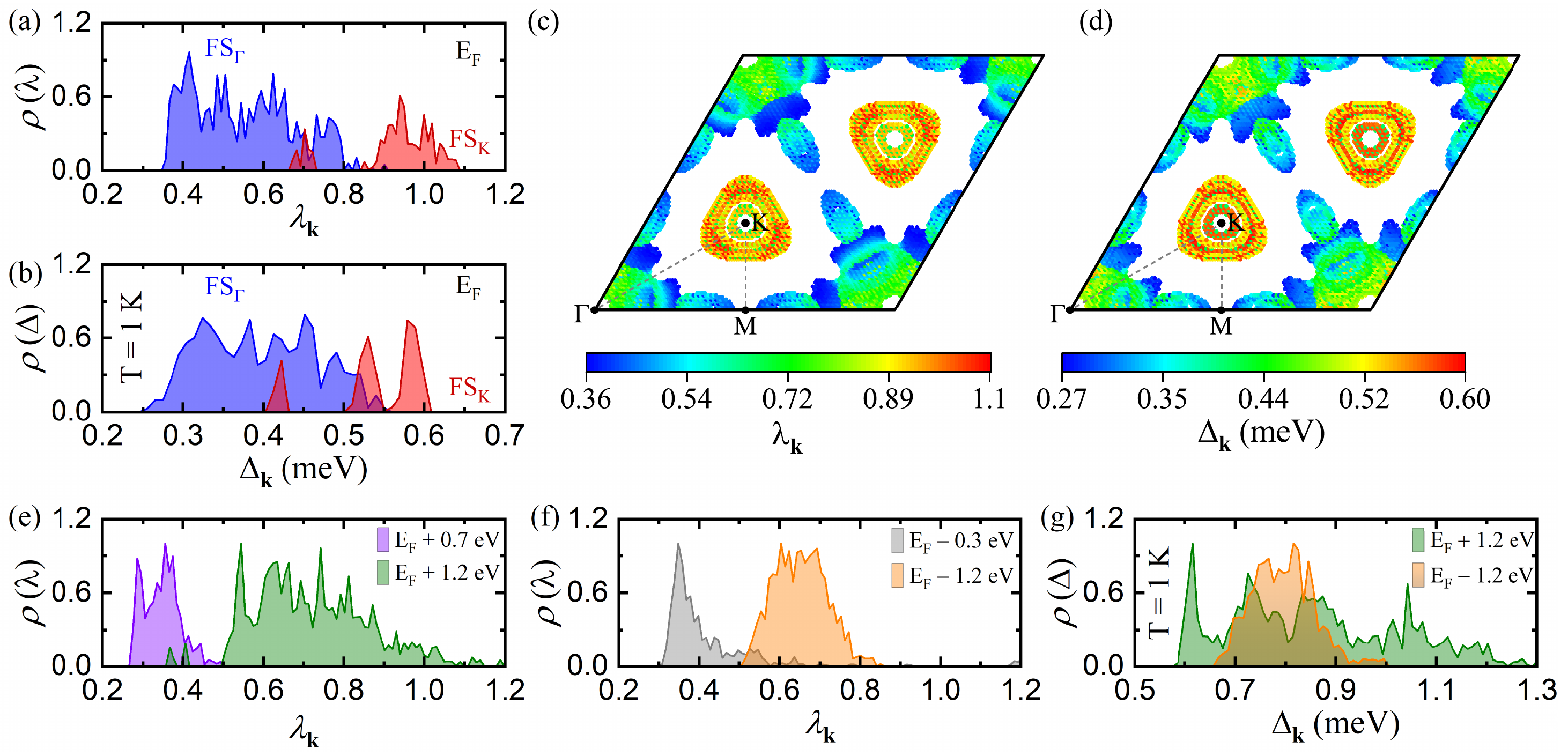}
	\caption{\label{fig:supercond_combine} Calculated superconducting properties for the $H$1 structure in the $\s1$ unit cell of SnSe$_2$ at the experimental unit cell parameters at 23~GPa. Energy distribution of the (a) e-ph coupling strength $\lambda_{\bf k}$ and (b) superconducting gap $\Delta_{\bf k}$; color coded by FS sheets: $\Gamma$-centered holelike pocket ${\rm FS}_\Gamma$ (blue), and $K$-centered electronlike pocket ${\rm FS}_K$ (red). Momentum-resolved (c) e-ph coupling strength $\lambda_{\bf k}$ and (d) superconducting gap $\Delta_{\bf k}$ on the FS (top-view). Energy distribution of the (e) e-ph coupling strength $\lambda_{\bf k}$, and (f), (g) superconducting gap $\Delta_{\bf k}$ at $T = 1$~K at various rigid shifts of the Fermi level with respect to the original data as shown in Supplemental Fig.~S14~\cite{SM}.}
\end{figure*}

Considering that superconductivity in SnSe$_2$ has been achieved experimentally either in Se-deficient~\cite{Zhou2018} or intercalated~\cite{Ohara1992,Song2019,Wu2019,Li2017} samples, we further investigate the impact of doping on the predicted $T_{\rm c}$ via a rigid shift of $E_{\rm F}$, chosen to match particular features either in the valence or in the conduction band, for the structure at the experimental lattice cell parameters at 23~GPa. The results summarized in Fig.~\ref{fig:supercond} can be rationalized in terms of the changes taking place at the FS upon raising or lowering the Fermi level. In the undoped systems, the momentum-resolved e-ph coupling strength $\lambda_{\bf k}$ on the FS  displays a highly anisotropic distribution (between 0.3 and 1.1) where the lower and upper regions of the spectra can be associated with the $\Gamma$ and $K$-centered pockets, respectively [Fig.~\ref{fig:supercond_combine}(a),(c)]. A continuous distribution with a sizable anisotropy is also found for the superconducting gap on the FS as shown in Fig.~\ref{fig:supercond_combine}(b),(d). When $E_{\rm F}$ is moved up by 0.7~eV, the spread in $\lambda_{\bf k}$ is reduced to a much narrower range of 0.3-0.5 since the $\Gamma$-centered holelike FS sheet vanishes. However, the spread goes back to the wider range once the system is doped until the Fermi level matches the first peak in the conduction band DOS as two additional FS sheets are introduced, thus opening extra scattering channels. This leads to an enhancement in both $\lambda=0.72$ and $T_{\rm c} \approx 5.2$~K, the latter in good agreement with the maximum $T_{\rm c}$ value found experimentally.  Similar trends are also observed when the Fermi level is shifted down. It is also worth noting that the non-linear dependence of the total $\lambda$ with doping correlates closely with that of the low-energy phonons [Fig.~\ref{fig:supercond}], indicating that these modes play a decisive role in raising the e-ph coupling strength and $T_{\rm c}$ in this system. 

For completeness, we also analyze the superconducting properties of SnS$_2$. Assuming that the dynamical stability of SnS$_2$ can be extended to higher pressures under nonhydrostatic compression as for SnSe$_2$, we perform superconductivity calculations at 30~GPa and used a smearing value of 0.04~Ry for the phonons at the $K$ point in order to remove the imaginary frequency. The calculated PHDOS and Eliashberg spectral function comprise of two well defined regions separated by a gap between the acoustic and optical branches [Supplemental Fig.~S18~\cite{SM}]. Integrating  $\alpha^2F(\omega)$, the strength of the e-ph coupling $\lambda$ is estimated to be 1.08. From the solutions of the isotropic and anisotropic ME gap equations with $\mu^* = 0.1$, we obtain a $T_{\rm c}$ of 4.6 and 5.9~K, respectively  [Supplemental Figs. S19-S20~\cite{SM}]. The values of $\lambda$ and $T_{\rm c}$ are twice larger than those in SnSe$_2$ due to the enhanced coupling of the acoustic phonons dominated by Sn vibrations with the electronic state present at the Fermi level. Similar to SnSe$_2$, the strongest coupling takes place on the $K$-centered FS, but both $\lambda_{\bf k}$ and $\Delta_{\bf k}$ exhibit a much broader distribution profile. We also find that $\lambda$ is strongly affected by changes in the FS topology when $E_{\rm F}$ is shifted up or down. Namely, the e-ph coupling drops significantly as the contribution of the low-energy acoustic modes is drastically reduced and large doping is required to raise back the $T_{\rm c}$. 

Finally, we investigate SnSe$_2$ and SnS$_2$ compounds in the $H$2-1 structure as potential superconductors. The results are presented in Supplemental Figs.~S21-S23~\cite{SM}. Despite the close resemblance between the DOS and PHDOS of the parent and superlattice structures, the $T_{\rm c}$ in $H$2-1 phase is found to be an order of magnitude lower than in the $H$1 phase. This can be linked to a decrease in the e-ph coupling over the full phonon range due to the reorganization that takes place at the FS. Looking at the electronic band structure in the $\sq3$ supercell [region enclosed by the green box in Fig.~\ref{fig:band_structures2}], it can be seen that the $\Gamma$-centered FS sheet disappears as a gap opens in the electronic band dominated by Se $p_z$ orbitals. An increase in the DOS at $E_{\rm F}$ is expected to produce a marked enhancement in $T_{\rm c}$, particularly under hole doping as multiple bands will cross the Fermi level. In SnSe$_2$ a  0.3~eV down shift in $E_{\rm F}$ results in an isotropic $T_{\rm c}=3.7$~K, while in SnS$_2$ a 0.5~eV down shift leads to $T_{\rm c}=3.5$~K. 
\
\section{Conclusions}
\label{sec:conclusions}

In this work, we have performed a comparative study of SnSe$_2$ and SnS$_2$ at ambient conditions and pressures up to 40~GPa. We show that the apparent contradictions among high-pressure results on SnSe$_2$ can be attributed to differences in experimental conditions and that inclusion of nonadiabatic effects improves the quantitative agreement with the measured Raman-active phonon frequencies. We further demonstrate that a periodic lattice transition, of a similar origin to the one observed in SnSe$_2$, also occurs in SnS$_2$ above 20~GPa. In addition, we examine the nature of the superconducting state recently observed in SnSe$_2$ under nonhydrostatic pressure, and provide evidence that the superconducting transition can be explained within a standard phonon-mediated mechanism. The emergence of superconductivity with a comparable critical temperature in SnS$_2$ under similar experimental conditions is also predicted.  Finally, we show that in the high pressure PLD $H$2-1 phase the $T_{\rm c}$ is reduced by an order of magnitude compared to the high-symmetry $H$1 phase in both systems, a fact that we attribute to a restructuring and suppression of large parts of the Fermi surface. 
 
\section*{Conflicts of interest}
There are no conflicts to declare.

\begin{acknowledgments} 
The authors thank J. Ying, V. V. Struzhkin, and F. Caruso for useful discussions. C. H. acknowledges support by the Austrian Science Fund (FWF) Project No. P32144-N36 and the VSC-3 of the Vienna University of Technology. G. P. K., H. P., and E. R. M. acknowledge support from the National Science Foundation (Award No. OAC–1740263). This work used Spiedie cluster at Binghamton University and Comet cluster at the San Diego Supercomputer Center through allocation TG-DMR180071. Comet is a dedicated  XSEDE cluster~\cite{XSEDE}, which is supported by National Science Foundation Grant No. ACI-1548562.
\end{acknowledgments}


\begin{thebibliography}{100}
\bibitem{note2}
{In the case of those pressures, where SnSe$_2$ is in the metallic phase, $\gmnvkqb \times \gmnvkqs$ has been approximated using the standard expression $|\gmnvkq|^2$ due to the fact that $\ve(\bq, \w= 0)$ diverges for metals.}
\end{thebibliography}
\end{document}


\title{Supplemental Material \\Electronic, vibrational, and electron-phonon coupling properties in SnSe$_2$ and SnS$_2$ under pressure}
	
\author{Gyanu Prasad Kafle}
\affiliation{Department of Physics, Applied Physics, and Astronomy, Binghamton University-SUNY, Binghamton, New York 13902, USA}
\author{Christoph Heil}
\affiliation{Institute of Theoretical and Computational Physics, Graz University of Technology, NAWI Graz, 8010 Graz, Austria}
\author{Hari Paudyal}
\affiliation{Department of Physics, Applied Physics, and Astronomy, Binghamton University-SUNY, Binghamton, New York 13902, USA}
\author{Elena R. Margine}
\email{rmargine@binghamton.edu}
\affiliation{Department of Physics, Applied Physics, and Astronomy, Binghamton University-SUNY, Binghamton, New York 13902, USA}
	
\maketitle

\begin{figure}[h]
	\centering
	\includegraphics[width=0.75\columnwidth]{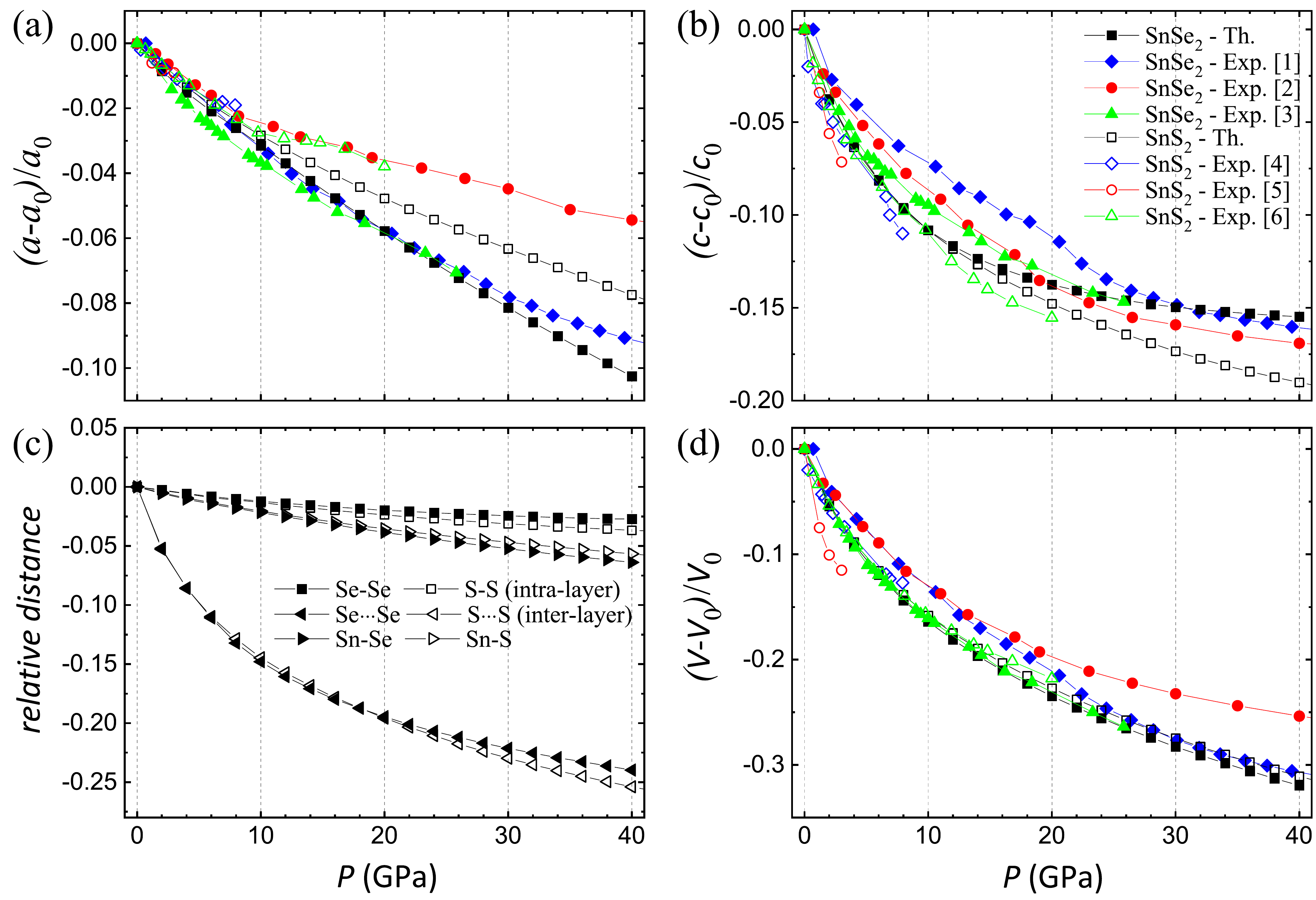}
	\caption{\label{figS1-lat-param-relative}  Pressure dependence of the relative change of (a)-(b) lattice parameters $a$ and $c$, (c) average bond lengths, and (d) volume per formula unit for SnSe$_2$ and SnS$_2$. Theoretical results are shown as black symbols and are compared with available experimental data~\cite{Zhou2018, Ying2018, Borges2018, Knorr2001, Hazen1978, Filso2016}.}
\end{figure}

\begin{figure}[h]
	\centering
	\includegraphics[width=0.75\columnwidth]{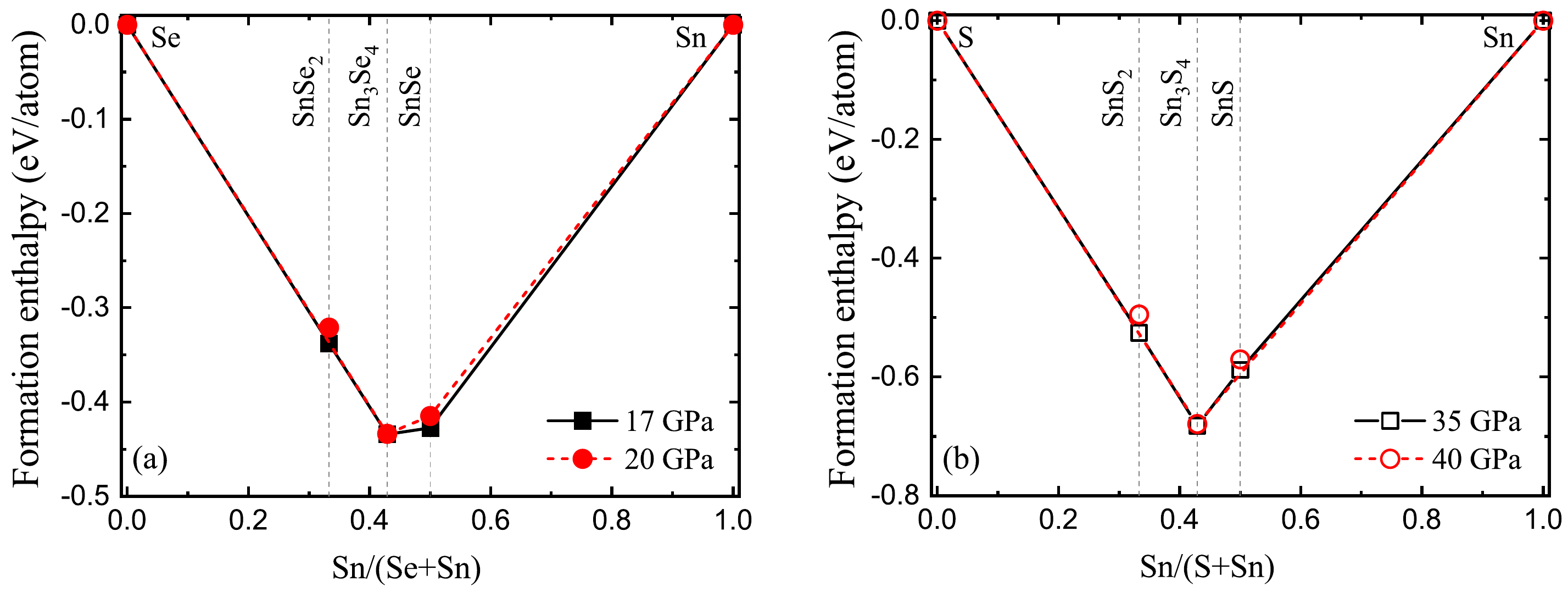}
	\caption{\label{figS2-convex-hull} Convex hulls of (a) Sn$_x$Se$_y$ and (b) Sn$_x$S$_y$ systems at various pressures.}
\end{figure}

\begin{figure}[h]
	\centering
	\includegraphics[width=0.9\columnwidth]{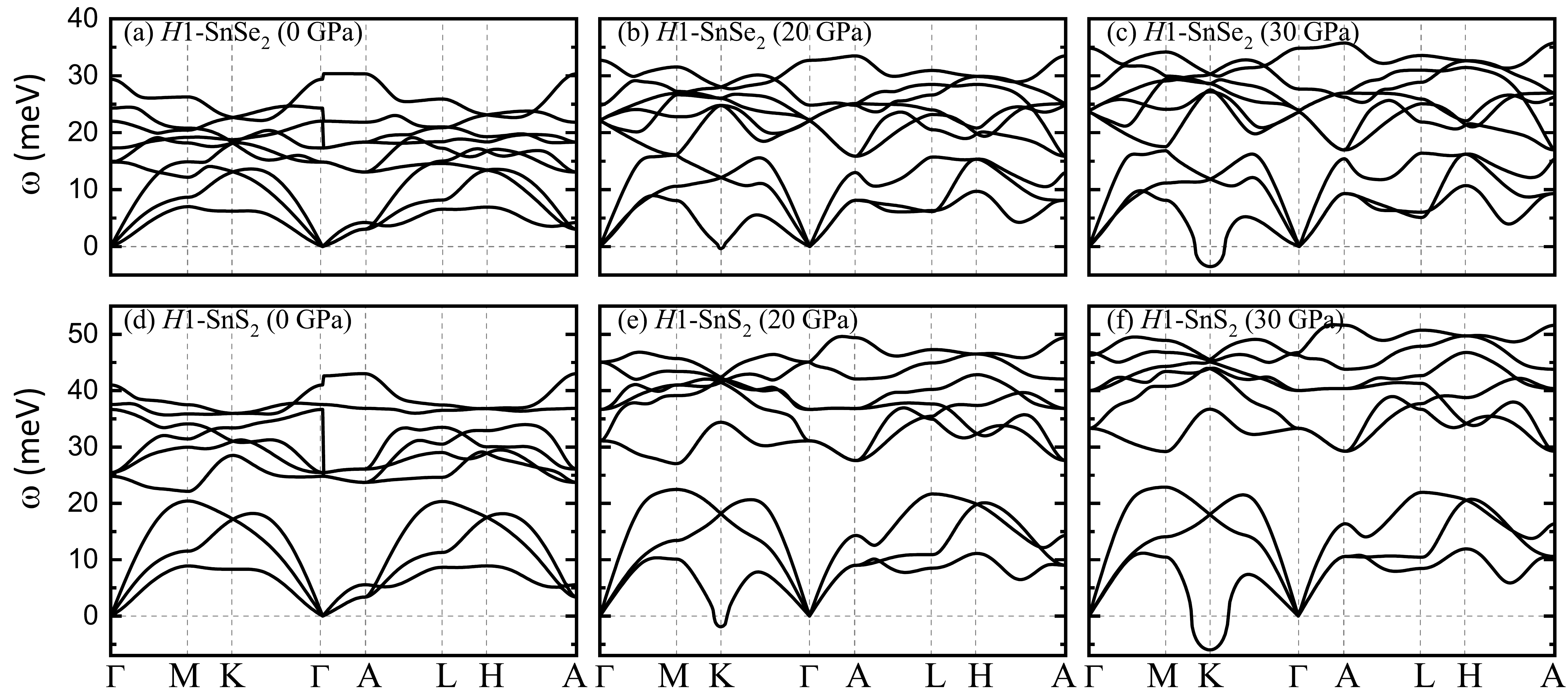}
	\caption{\label{figS3-ph-h1} Calculated phonon dispersion for the $H$1 structure in the $\s1$ unit cell of (a)-(c) SnSe$_2$ and (d)-(f) SnS$_2$ at 0, 20, and 30 GPa.}
\end{figure}

\begin{table}[h!]
	\vspace{0.3cm} 
	\centering
	\resizebox{!}{0.22\textwidth}{%
		\begin{tabular}{|c|c|c|c|c|c|c|c|c c|}
			\hline
			\textbf{System}  & \textbf{Phase}  & \textbf{Space} & \textbf{Eigenvector} & \textbf{$\Delta H$} & \textbf{Similarity} & \textbf{$a$}  & \textbf{$c$}  & \multicolumn{2}{|c|}{\textbf{Wyckoff positions}}\\
			&  &  \textbf{group} & \textbf{combination} & \textbf{(meV/atom)} & \textbf{factor} & ({\textbf{\AA}}) & ({\textbf{\AA}}) & \textbf{Sn} & \textbf{Se/S} \\
			\hline
			\multirow{2}{*}{\textbf{SnSe$_2$}}  & $H$1  & 164 &  &  &  & 3.86169 & 6.10587 & 1a (0.00000, 0.00000, 0.00000) & 2d     (0.33333, 0.66667, 0.23363)\\ 
			& (0 GPa) & P$\bar{3}$m1 &  &  &  &  &  &  & \\ \cline{2-10}
			& $H$1 & 164 &  &  &  & 3.54726 & 5.19218 & 1a (0.00000, 0.00000, 0.00000) & 2d     (0.33333, 0.66667, 0.30816)\\
			& (30 GPa) &  P$\bar{3}$m1 &  &  &  &  &  &  & \\ \cline{2-10}
			& $H$2-1  & 147 & $e_1$ & -1.86477 & 0.4983 &6.14654 & 5.17350 & 2d (0.33333, 0.66667, 0.95799) & 6g (0.02978, 0.34880, 0.30584)\\
			& (30 GPa) &  P$\bar{3}$ &  &  &  &  &  & 1a (0.00000, 0.00000, 0.00000) & \\ \cline{2-10}
			& $H$2-2  & 157 & $e_2$ & -1.83621 & 0.2058 & 6.14662 & 5.17334 & 2b (0.33333, 0.66667, 0.97571) &  3c (0.00000, 0.35864, 0.30583)  \\
			& (30 GPa) &  P31m &  &  &  &  &  & 1a (0.00000, 0.00000, 0.04855) & 3c (0.00000, 0.69306, 0.69416) \\ \cline{2-10}
			\hline
			\multirow{2}{*}{\textbf{SnS$_2$}}  & $H$1  & 164 &  &  &  & 3.68056 & 5.88359 & 1a (0.00000, 0.00000, 0.00000) & 2d     (0.33333, 0.66667, 0.25197)\\ 
			& (0 GPa) & P$\bar{3}$m1 &  &  &  &  &  &  & \\ \cline{2-10}
			& $H$1 & 164 &  &  &  & 3.45051 & 4.84234 & 1a (0.00000, 0.00000, 0.00000) & 2d     (0.33333, 0.66667, 0.30146)\\
			& (30 GPa) &  P$\bar{3}$m1 &  &  &  &  &  &  & \\ \cline{2-10}
			& $H$2-1  & 147 & $e_1$ & -1.30561 & 0.6158 & 5.96638 & 4.84894 & 2d (0.33333, 0.66667, 0.03189) & 6g (0.97915, 0.32334, 0.30017)\\
			& (30 GPa) &  P$\bar{3}$ &  &  &  &  &  & 1a (0.00000, 0.00000, 0.00000) & \\ \cline{2-10}
			& $H$2-2  & 157 & $e_2$ & -1.31027 & 0.4223 & 5.96619 & 4.84884 & 2b (0.33333, 0.66667, 0.01870) &  3c (0.00000, 0.31459, 0.30014)  \\
			& (30 GPa) &  P31m &  &  &  &  &  & 1a (0.00000, 0.00000, 0.96262) & 3c (0.00000, 0.64879, 0.69986) \\ \cline{2-10}
			& $H$2-3 & 143 & $e_1 + e_2$ & -1.32715 & 0.5887 & 5.97630 & 4.84268 & 1a (0.00000, 0.00000, 0.97419) &  3d (0.99372, 0.31659, 0.30144)  \\
			& (30 GPa)&  P3 &  &  &  &  &  & 1b (0.33333, 0.66667, 0.02175) & 3d (0.00628, 0.65620, 0.69856) \\
			&  &  &  &  &  &  &  & 1c (0.66667, 0.33333, 0.00407) &  \\
			\hline
		\end{tabular}
	}
	\caption{\label{table1}
	Space groups, lattice parameters, and Wyckoff positions of SnSe$_2$ and SnS$_2$ structures fully relaxed at the DFT level. The space groups were found with the SPGLIB tool~\cite{SPGLIB} interfaced with the MAISE package~\cite{Hajinazar2020} using a tolerance of 0.01. The lattice parameters are given  for $H$1 at 0~GPa and 30~GPa, and  for the $H$2 derivatives at 30~GPa. The $H$2 derivatives were constructed from the $\sq3$ $H$1 supercell by considering nonequivalent combinations of the $e_1$ and $e_2$ eigenvectors corresponding to the two nearly degenerate lowest-energy phonon modes $A_{2g}$ and $A_{2u}$. Configuration $H$2-1 corresponds to atomic displacements along $e_1$, configuration $H$2-2  corresponds to atomic displacements along $e_2$, and configuration $H$2-3 corresponds to atomic displacements along a linear combination of the two eigenvectors. The similarity factor was calculated for each $H$2 derivative  with respect to the undistorted $\sq3$ $H$1 supercell.}

\end{table}

\begin{figure}[h]
	\centering
	\includegraphics[width=0.75\columnwidth]{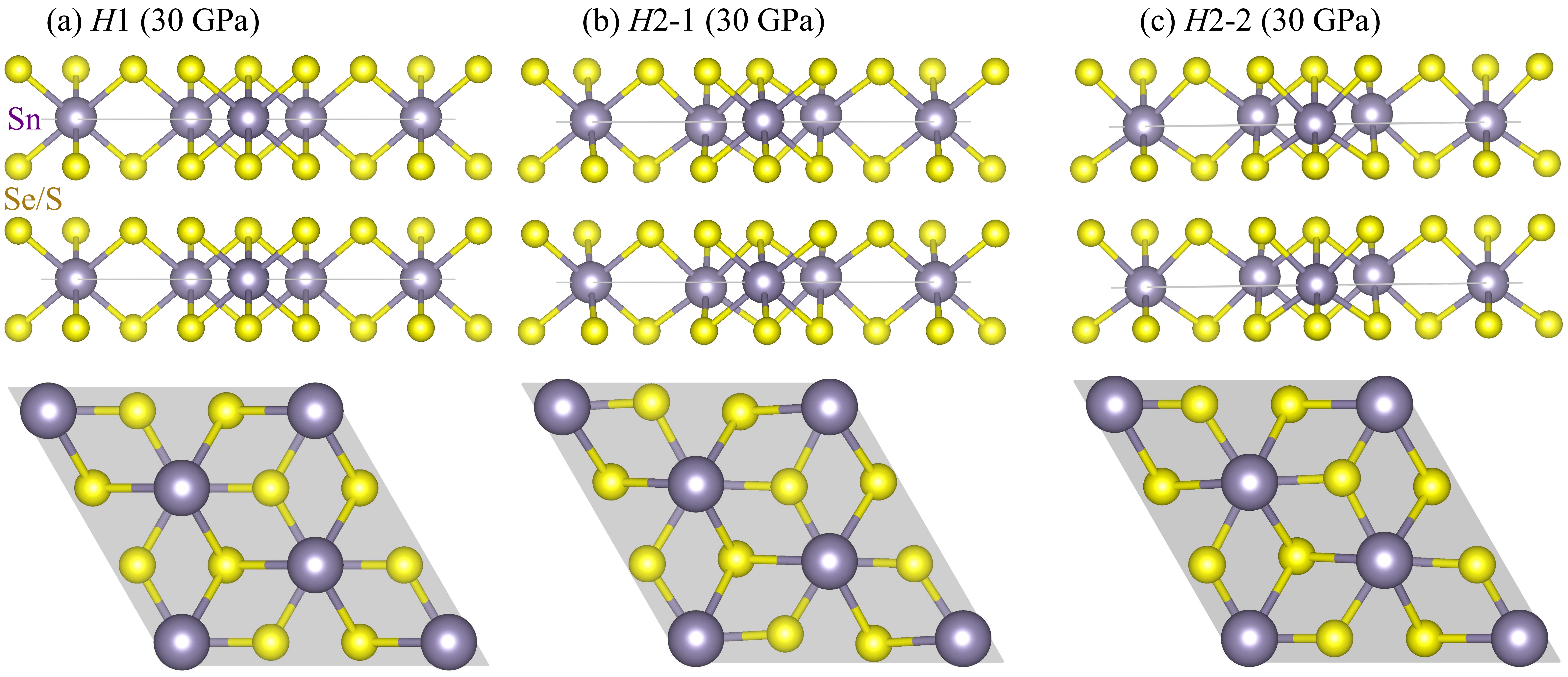}
	\caption{\label{figS4-crystal-structure} Crystal structures (side and top view) for the (a) $H$1, (b) $H$2-1, and (c) $H$2-2 structures in the $\sq3$ supercell of SnSe$_2$ and SnS$_2$, fully relaxed with the DFT at 30 GPa. }
\end{figure}

\begin{figure}[h]
	\centering
	\includegraphics[width=0.9\columnwidth]{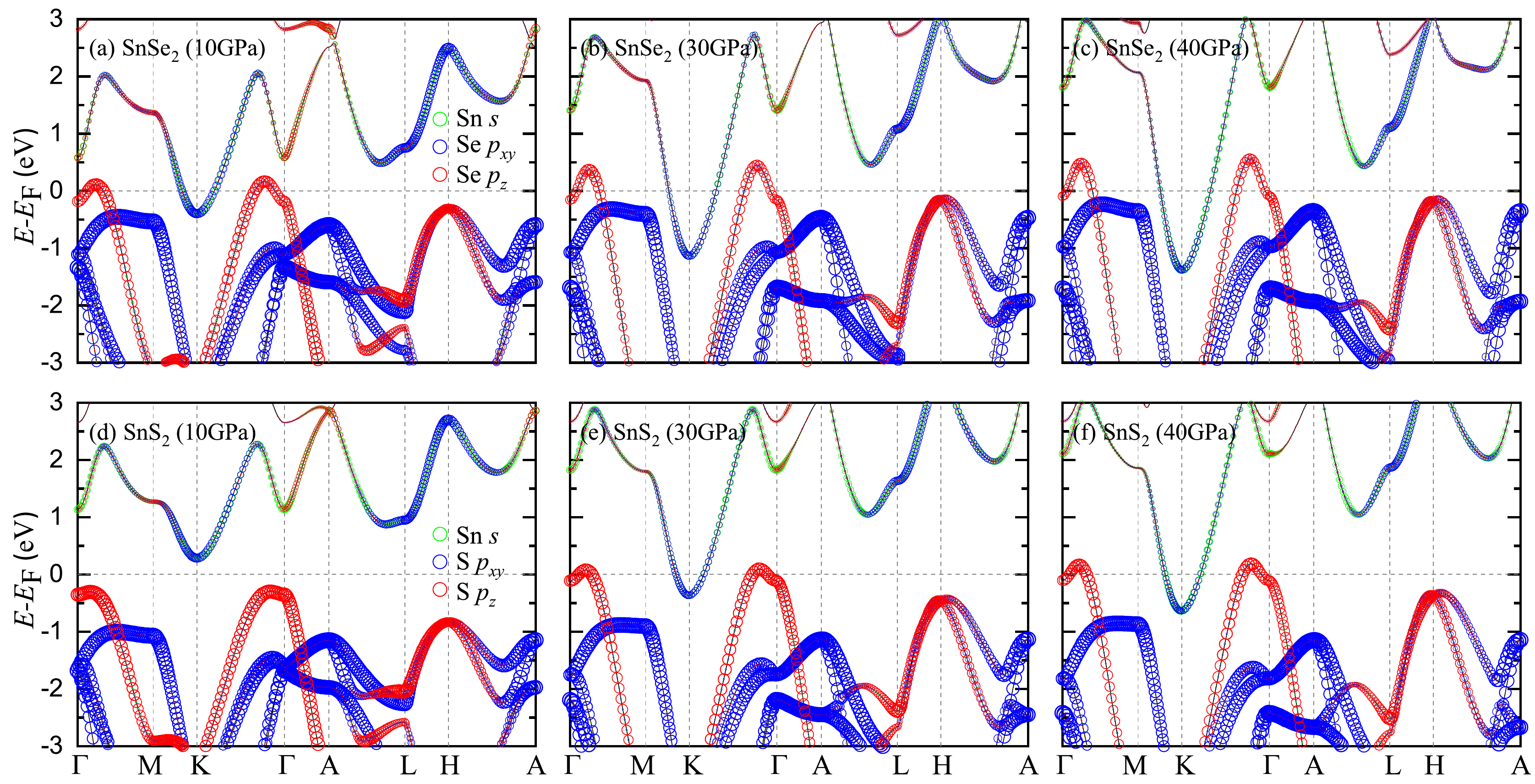}
	\caption{\label{figS5-char-band} Calculated band structure  for the $H$1 structure in the $\s1$ unit cell of (a)-(c) SnSe$_2$ and (d)-(f) SnS$_2$ at 10, 30, and 40~GPa. The size of the symbols is proportional to the contribution of each orbital character.}
\end{figure}

\begin{figure}[h]
	\centering
	\includegraphics[width=0.60\columnwidth]{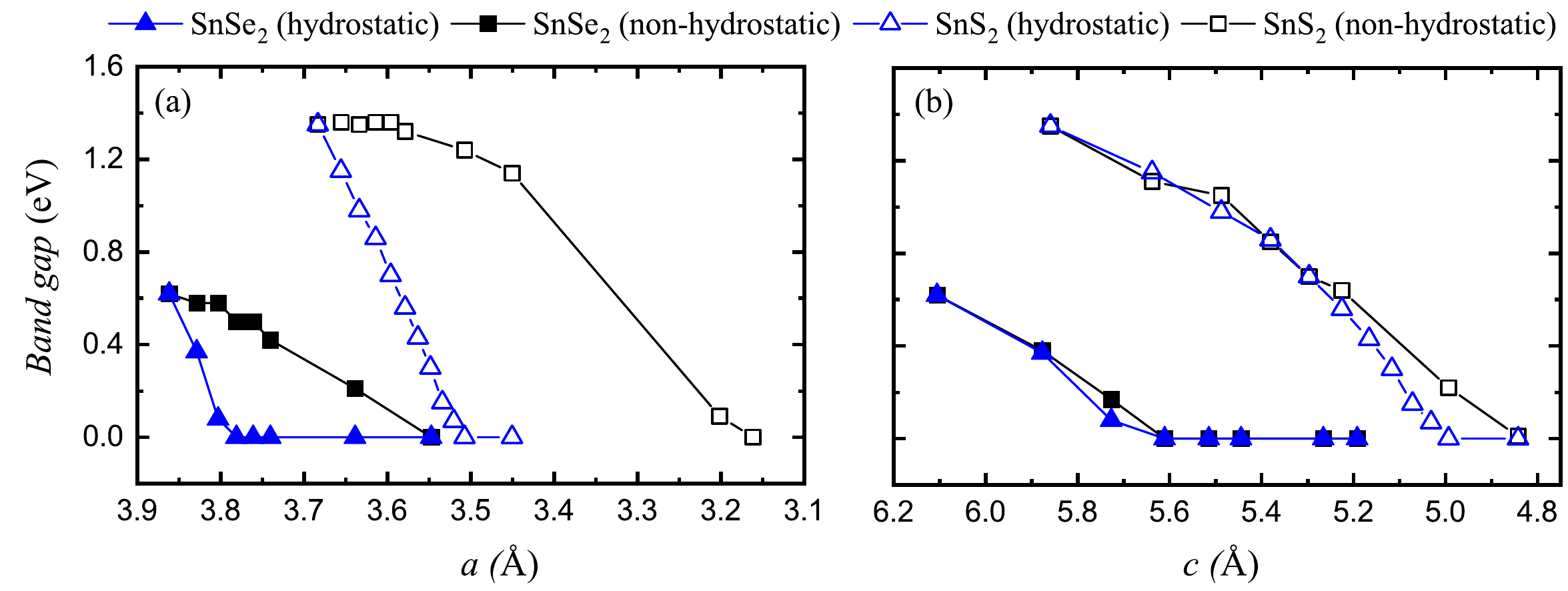}
	\caption{\label{figS6-gap-lat-nonhydro} Calculated band gap for the $H$1 structure in the $\s1$ unit cell of SnSe$_2$ and SnS$_2$ under hydrostatic and non-hydrostatic pressure as a function of lattice parameters (a) $a$ and (b) $c$. In (a), $c$ is kept fixed at $c$ = 6.10~\AA\, for SnSe$_2$ and $c$ = 5.86~\AA\, for SnS$_2$. In (b), $a$ is kept fixed at $a$ = 3.86~\AA\, for SnSe$_2$ and $a$ = 3.68~\AA\, for SnS$_2$.}
\end{figure}

\begin{figure}[h]
	\centering
	\includegraphics[width=0.70\columnwidth]{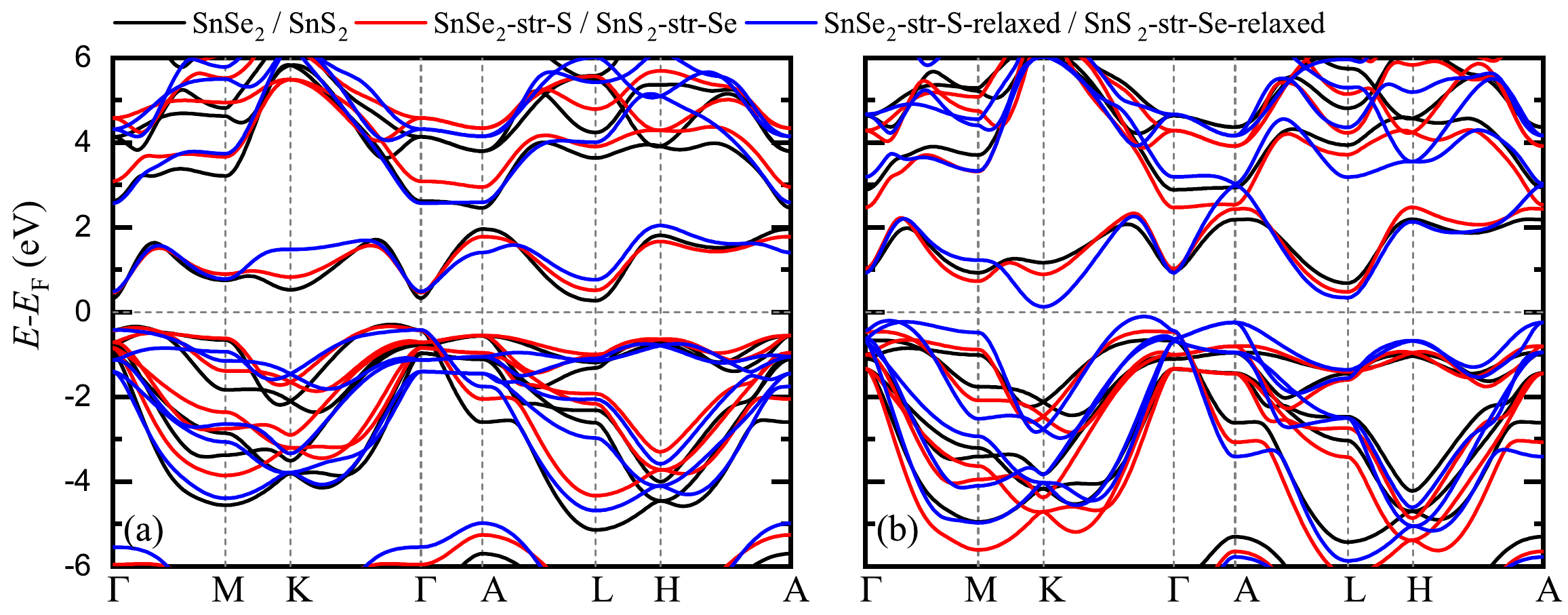}
	\caption{\label{figS7-band0}  Calculated band structure  for the $H$1 structure in the $\s1$ unit cell  of (a) SnSe$_2$, SnSe$_2$-str-S, and SnSe$_2$-str-S-relaxed and (b) SnS$_2$, SnS$_2$-str-Se, and SnS$_2$-str-Se-relaxed at 0 GPa.} 
\end{figure}

\begin{figure}[h]
	\centering
	\includegraphics[width=0.85\linewidth]{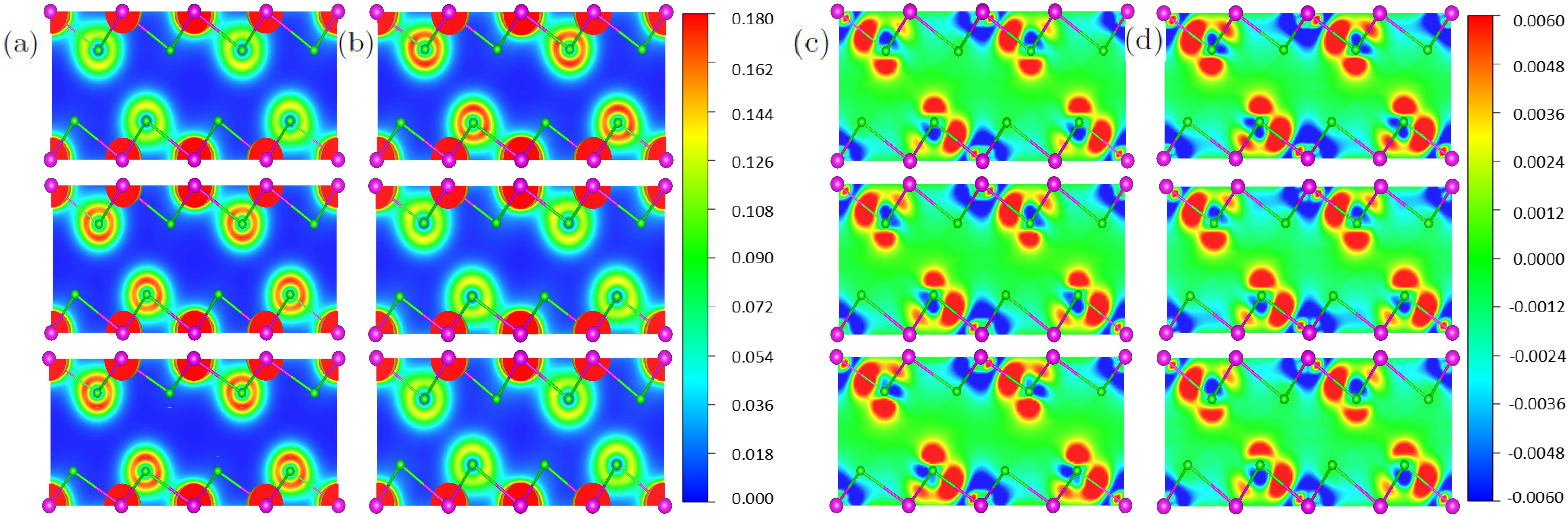}
	\caption{\label{figS9-charge0-vesta} Calculated (a) charge density and (c) charge density difference along the (110) plane for the $H$1 structure in the $\s1$ unit cell of SnSe$_2$, SnSe$_2$-str-S, and SnSe$_2$-str-S-relaxed (from top to bottom)  at 0~GPa. Same for SnS$_2$, SnS$_2$-str-Se, and SnS$_2$-str-Se-relaxed in (b) and (d).} 
\end{figure}

\begin{figure}[h]
	\centering
	\includegraphics[width=0.6\linewidth]{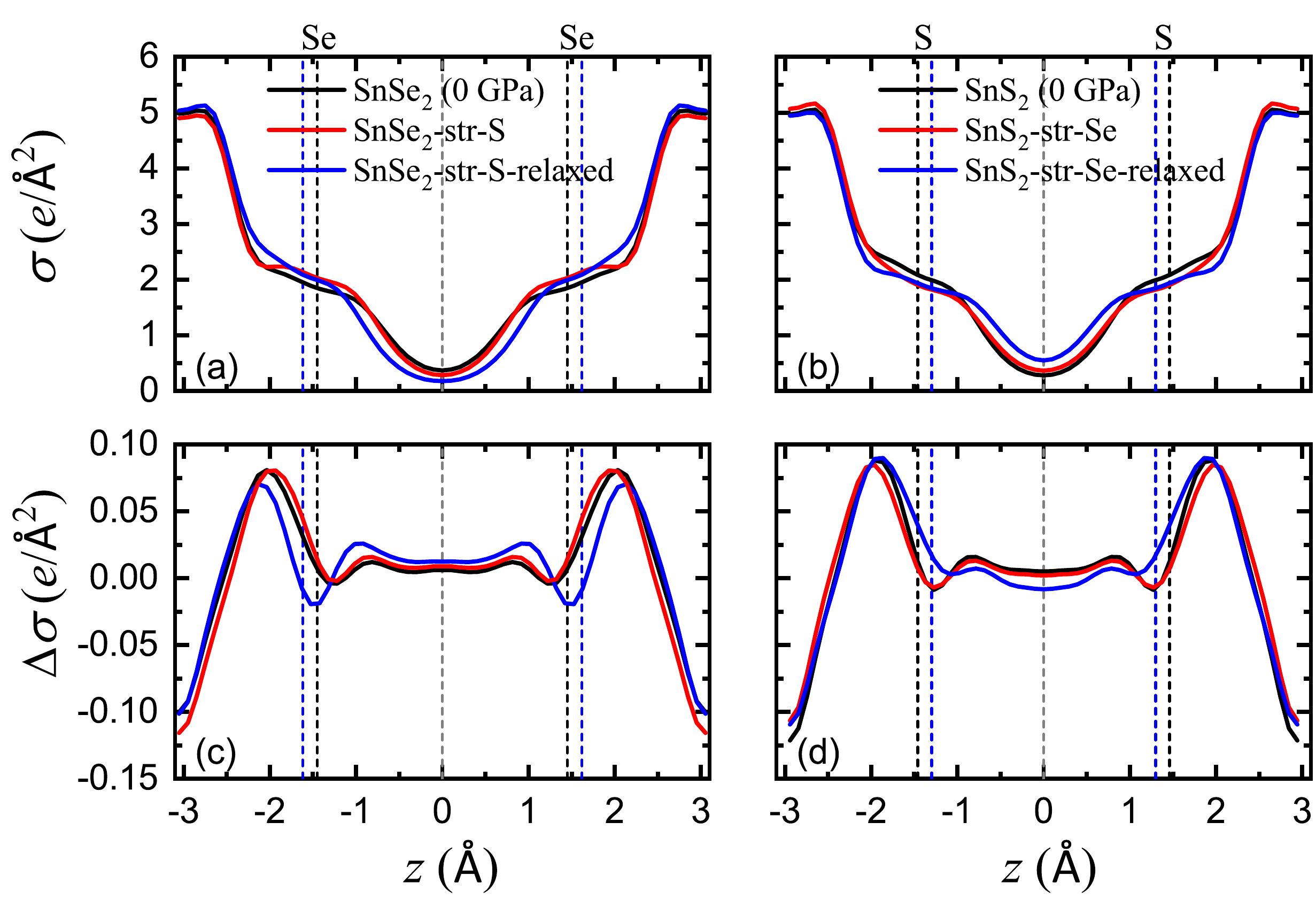}
	\caption{\label{figS8-charge0} Calculated (a) charge density and (c) charge density difference (in the $x$-$y$ plane) as a function of the perpendicular direction $z$ for the $H$1 structure in the $\s1$ unit cell of SnSe$_2$, SnSe$_2$-str-S, and SnSe$_2$-str-S-relaxed at 0 GPa. Same for SnS$_2$, SnS$_2$-str-Se, and SnS$_2$-str-Se-relaxed in (b) and (d). The vertical black and blue dashed lines represent the position of the Se/S atoms  for SnSe$_2$/SnS$_2$ and SnSe$_2$-str-S-relaxed/SnS$_2$-str-Se-relaxed structures along $c$-axis, and the vertical  gray dashed line represents the middle of van der Waals gap.} 
\end{figure}

\begin{figure}[h]
	\centering
	\includegraphics[width=0.85\linewidth]{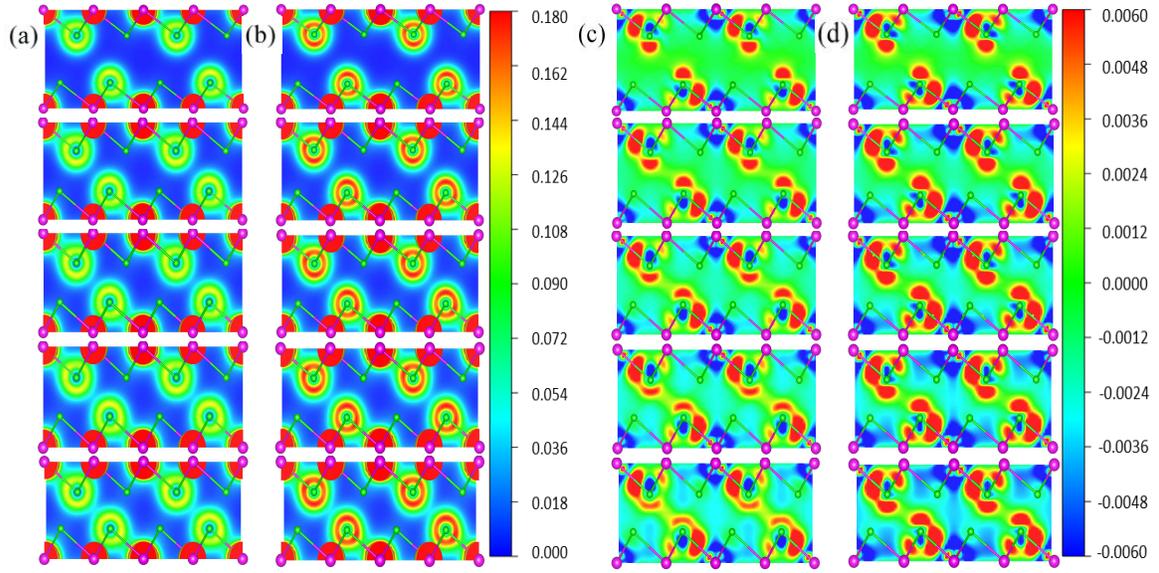}
	\caption{\label{figS11-charge-dens-vesta} Calculated (a) charge density and (c) charge density difference along the (110) plane for the $H$1 structure in the $\s1$ unit cell of SnSe$_2$ at various pressures. Same for SnS$_2$ in (b) and (d). The pressure order  is 0, 10, 20, 30, and 40~GPa from top to bottom.} 
\end{figure}

\begin{figure}[h]
	\centering
	\includegraphics[width=0.6\linewidth]{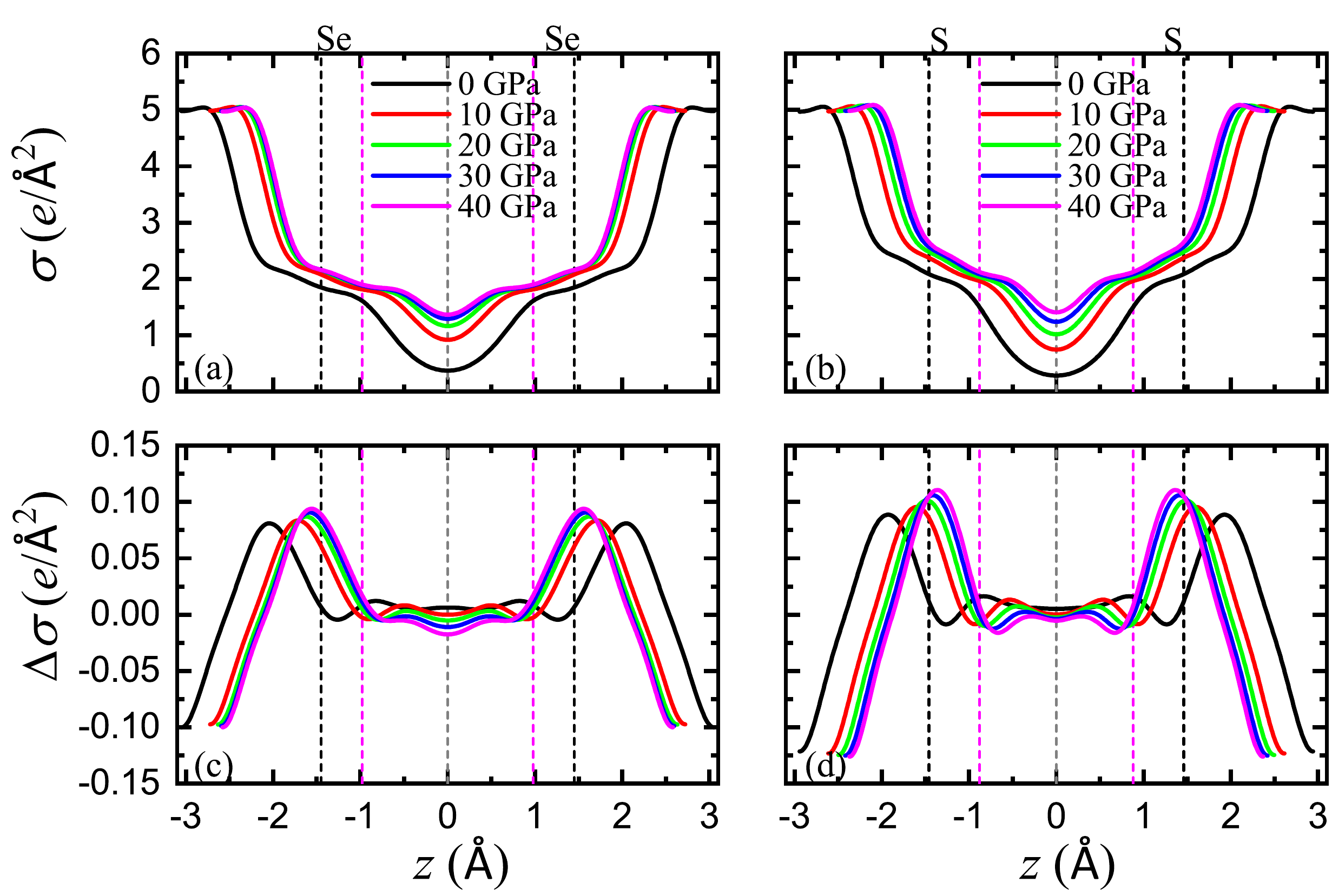}
	\caption{\label{figS10-charge-dens} Calculated (a) charge density and (c) charge density difference (in the $x$-$y$ plane) as a function of the perpendicular direction $z$ for the $H$1 structure in the $\s1$ unit cell of SnSe$_2$ at various pressures. Same for SnS$_2$ in (b) and (d). The vertical black and magenta dashed lines represent the position of the Se/S atoms  for SnSe$_2$/SnS$_2$ structures along $c$-axis at 0 and 40 GPa, respectively, and the vertical  gray dashed line represents the middle of van der Waals gap.} 
\end{figure}


\begin{figure}[h]
	\centering
	\includegraphics[width=0.4\columnwidth]{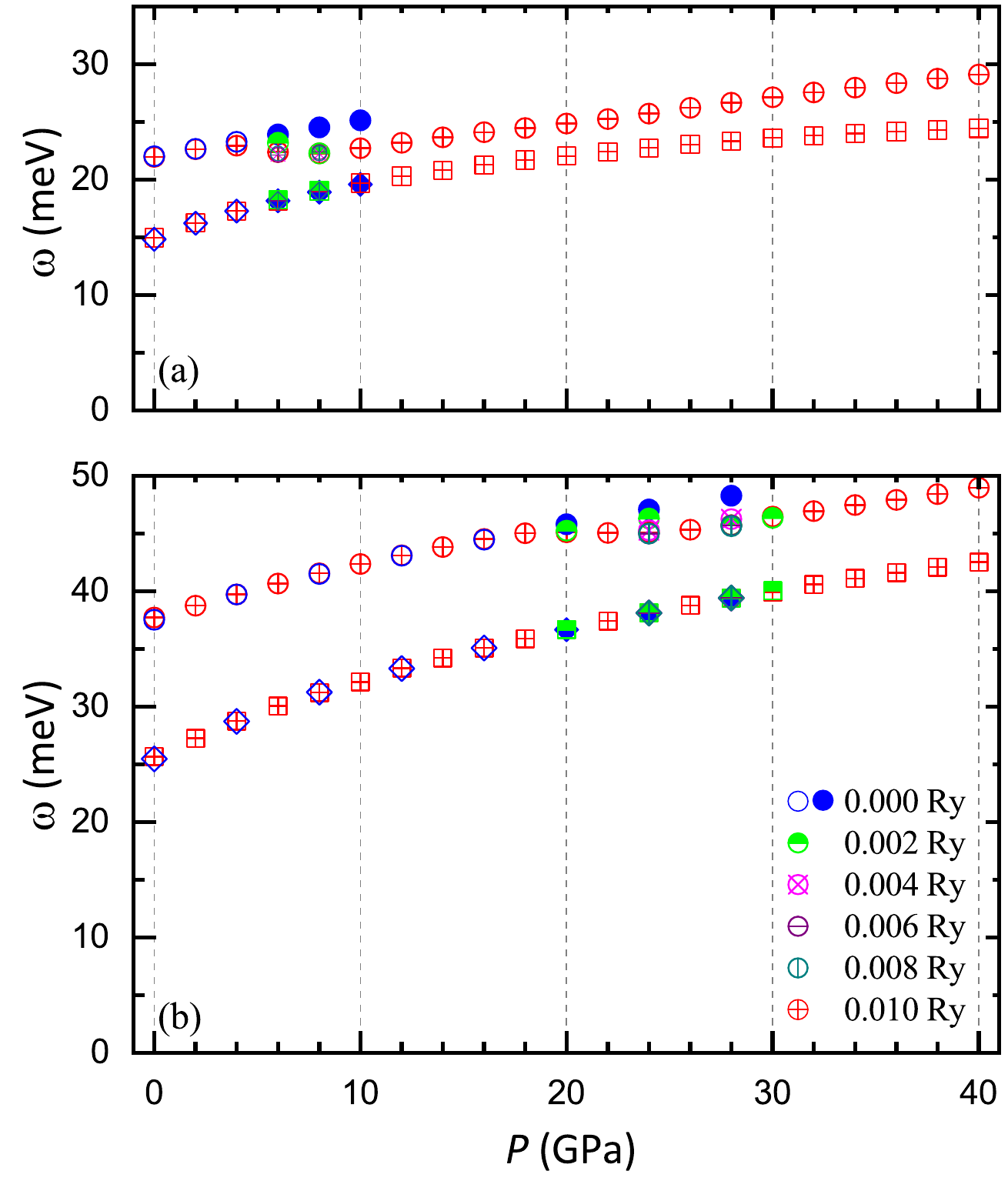}
	\caption{\label{figS12-raman} Calculated frequency dependence of the Raman-active modes, $A_{1g}$ and $E_{g}$,  for the $H$1 structure in the $\s1$ unit cell of (a) SnSe$_2$ and (b) SnS$_2$ as a function of pressure using various smearing values. The data before and after the metalization (6 GPa in SnSe$_2$ and 20 GPa in SnS$_2$) using no smearing are shown as open and filled blue circles, respectively.}
\end{figure}

\begin{figure}[h]
	\centering
	\includegraphics[width=0.9\columnwidth]{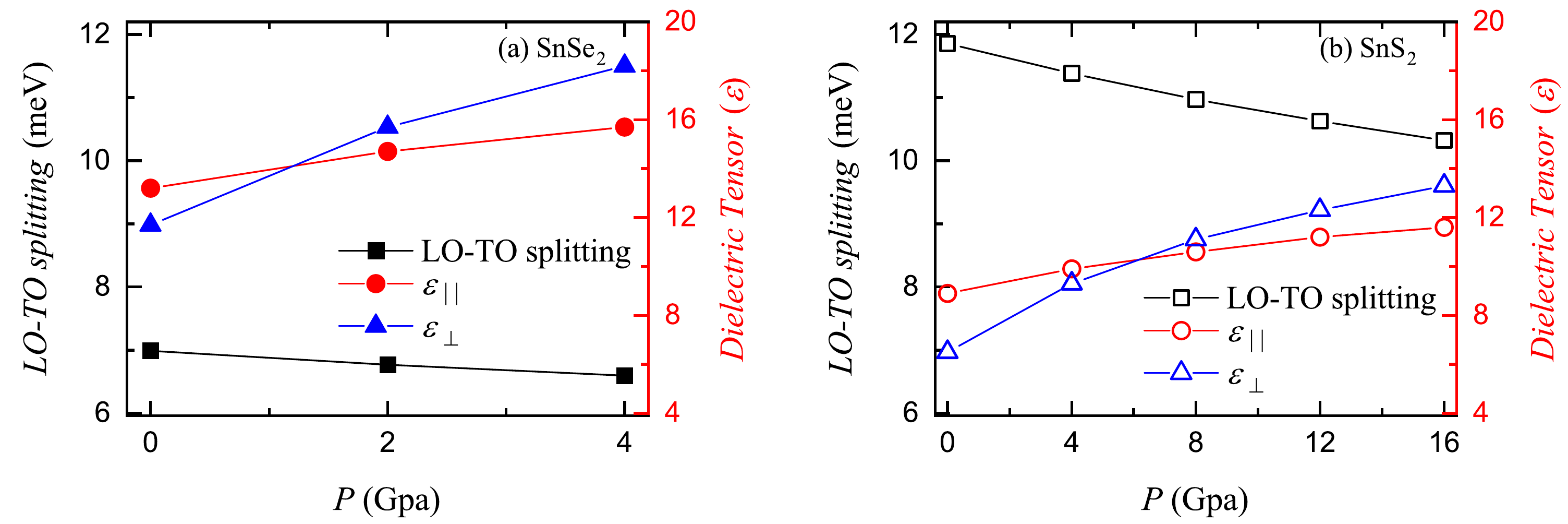}
	\caption{\label{figS13-LOTO} Calculated LO-TO splitting, and in-plane ($\varepsilon_{\|}$) and out-of-plane ($\varepsilon_{\perp}$) dielectric tensors for the $H$1 structure in the $\s1$ unit cell of (a) SnSe$_2$ and (b) SnS$_2$ as a function of pressure.} 
\end{figure}

\clearpage
\newpage
\begin{figure}[h]
	\centering
	\includegraphics[height=4.54 cm]{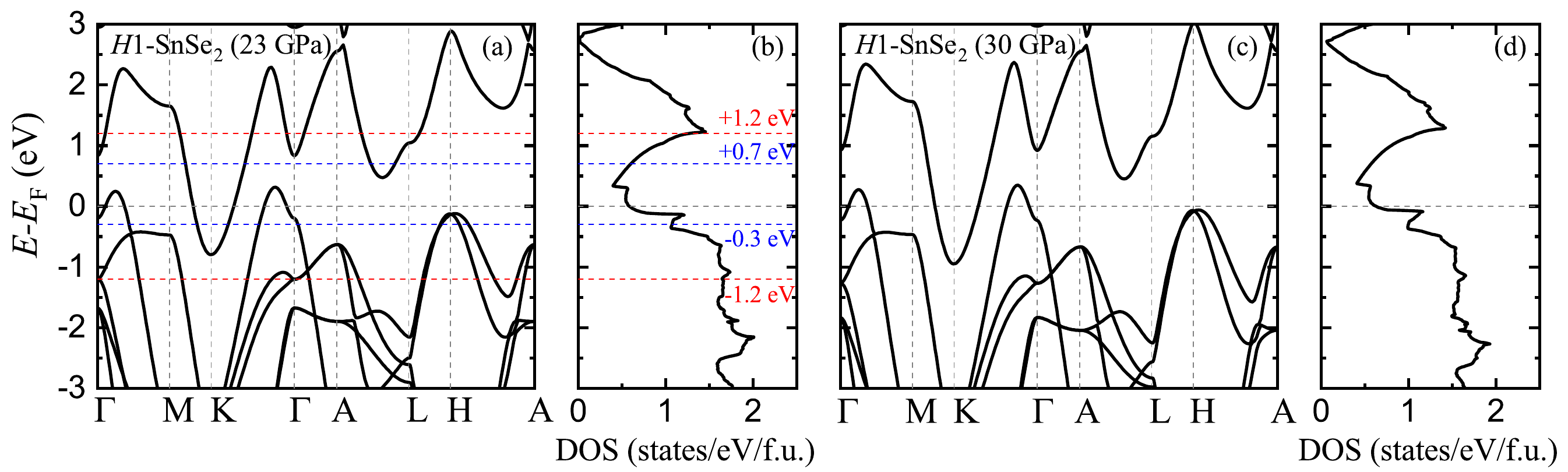}
	\caption{\label{figS14-SnSe2_h1-band-dos-23_30GPa-Zhou-expt} Calculated band structure and DOS for the $H$1 structure in the $\s1$ unit cell of SnSe$_2$ at the experimental unit cell parameters at (a)-(b) 23 and (c)-(d) 30~GPa. The red and blue dashed lines represent rigid shifts of the Fermi level with respect to the original data.} 
\end{figure}

\begin{figure}[h]
	\centering
	\includegraphics[height=4.54 cm]{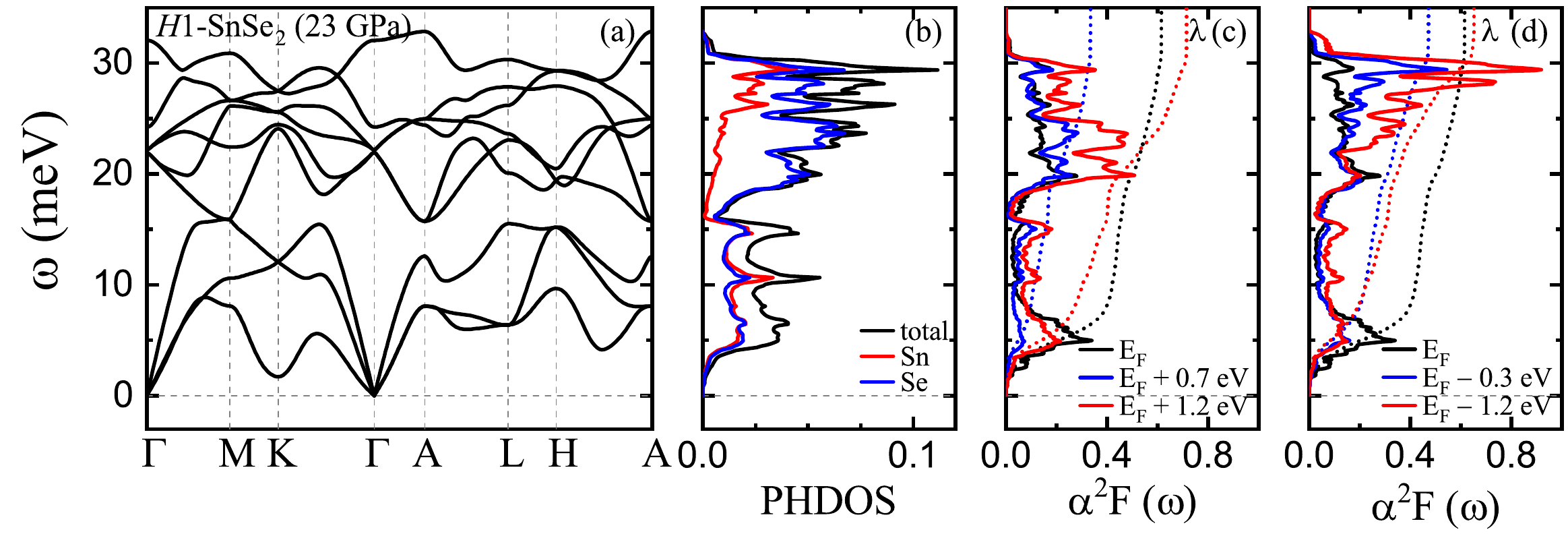}
	\caption{\label{figS15-SnSe2_h1-ph_phdos_a2f-23GPa-Zhou-et-al-expt} Calculated (a) phonon dispersion, (b) PHDOS, and (c)-(d) Eliashberg spectral function $\alpha^2F(\omega)$ and e-ph coupling strength $\lambda(\omega)$ for the $H$1 structure in the $\s1$ unit cell of SnSe$_2$ at the experimental lattice parameters at 23 GPa. In (c) and (d), black lines show $\alpha^2F(\omega)$ and $\lambda(\omega)$ at the Fermi level, while red and blue lines show the same quantities for the rigid shifts in the Fermi level indicated in Fig.~\ref{figS14-SnSe2_h1-band-dos-23_30GPa-Zhou-expt}(a)-(b).}
\end{figure}

\begin{figure}[h]
	\centering
	\includegraphics[height=4.54 cm]{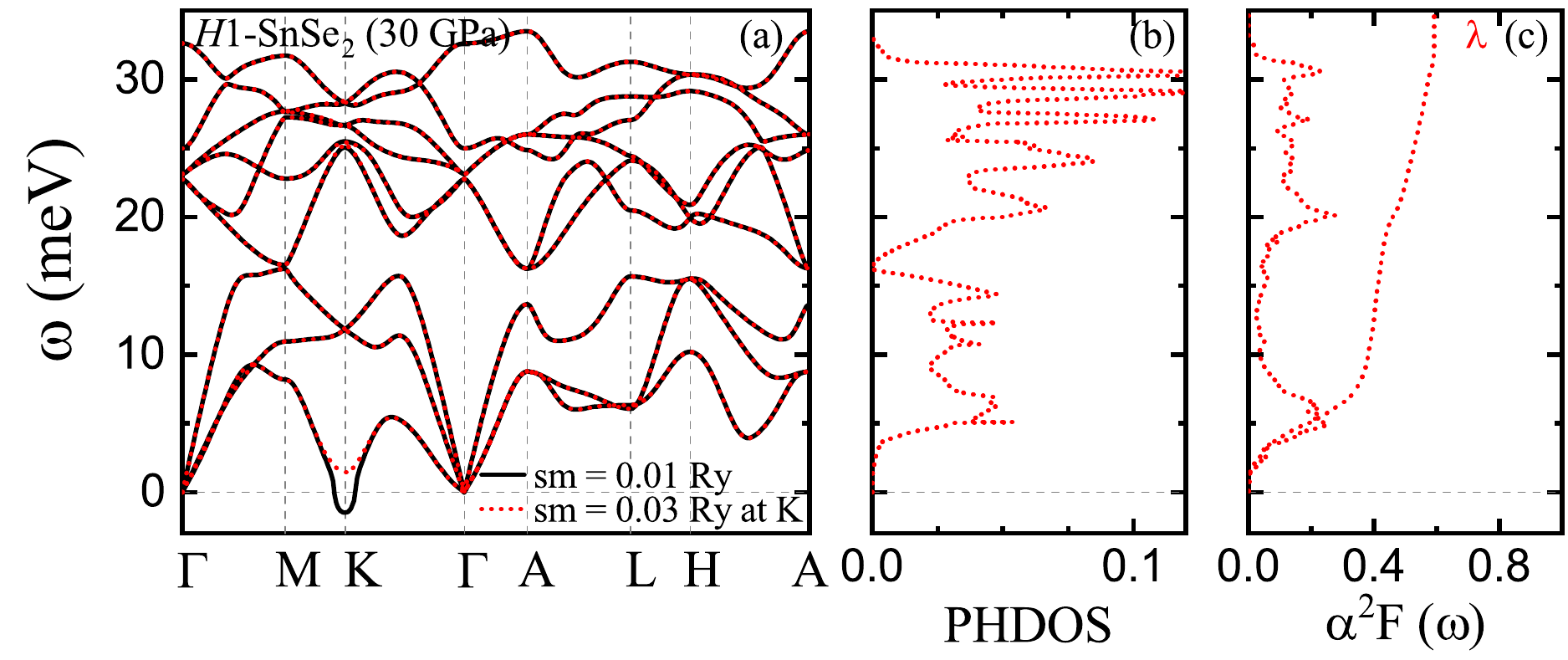}
	\caption{\label{figS16-SnSe2_h1-ph_phdos_a2f-30GPa-Zhou-et-al-expt} Calculated (a) phonon dispersion, (b) PHDOS, and (c) Eliashberg spectral function $\alpha^2F(\omega)$, and e-ph coupling strength $\lambda(\omega)$ for the $H$1 structure in the $\s1$ unit cell of SnSe$_2$ at the experimental lattice parameters at 30~GPa. The black lines represent the phonon calculated with a smearing value of 0.01~Ry, while red lines with a smearing value of 0.03~Ry at the $K$-point in the BZ.}
\end{figure}

\begin{figure}[h]
	\centering
	\includegraphics[height=4.78 cm]{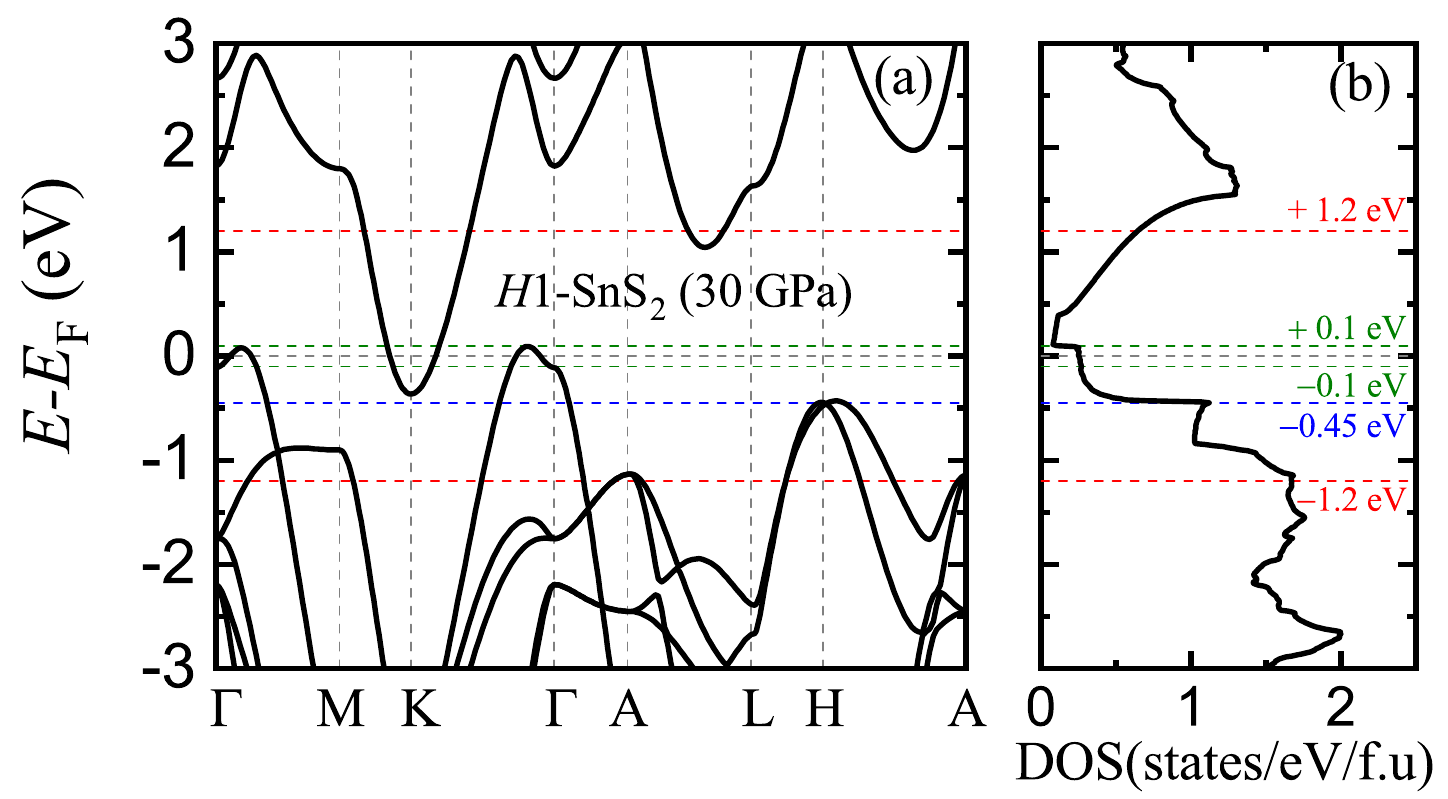}
	\caption{\label{figS17-SnS2_h1-band-dos-p30} Calculated (a) band structure and (b) DOS for the $H$1 structure in the $\s1$ unit cell of SnS$_2$ at 30~GPa. The red and blue dashed lines represent rigid shifts of the Fermi level with respect to the original data.}
\end{figure}

\begin{figure}[h]
	\centering
	\includegraphics[height=4.78 cm]{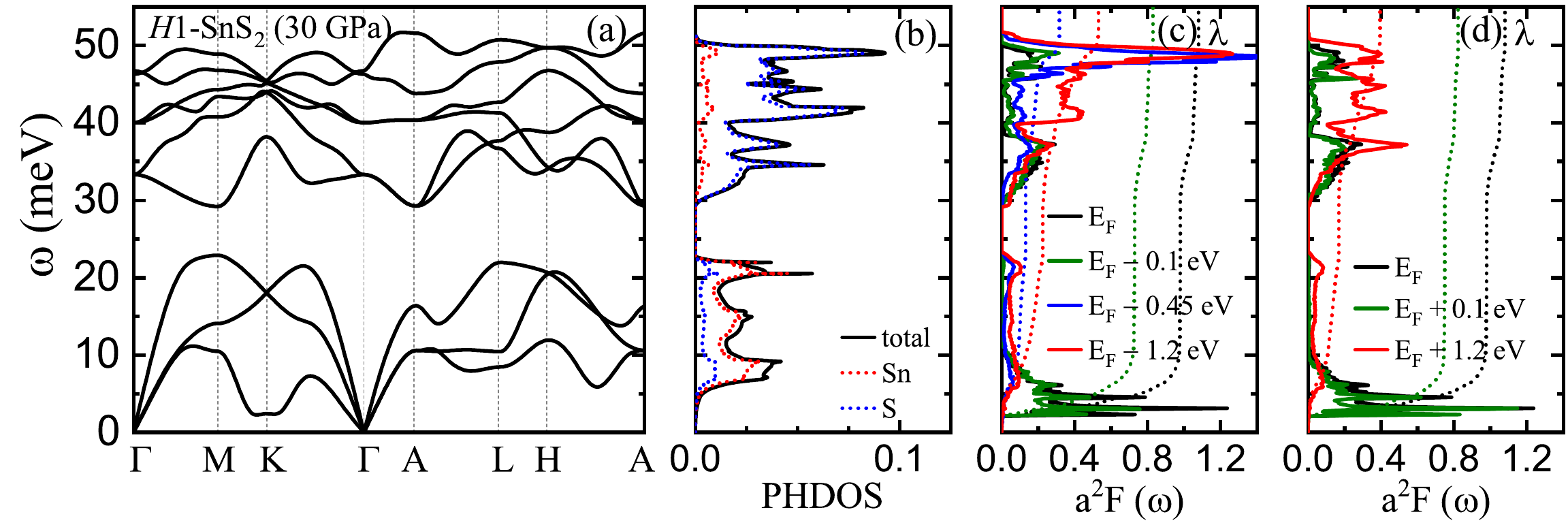}
	\caption{\label{figS18-SnS2_h1-ph-a2f-p30} Calculated (a) phonon dispersion, (b) PHDOS, and (c)-(d) Eliashberg spectral function $\alpha^2F(\omega)$, and e-ph coupling strength $\lambda(\omega)$ for the $H$1 structure in the $\s1$ unit cell of SnS$_2$ at 30 GPa. In (c) and (d), black lines show $\alpha^2F(\omega)$ and $\lambda(\omega)$ at the Fermi level, while red, blue and green lines show the same quantities for the rigid shifts in the Fermi level indicated in Fig.~\ref{figS17-SnS2_h1-band-dos-p30}. The phonon is calculated with a smearing value of 0.04~Ry at the $K$-point in the BZ.} 
\end{figure}

\begin{figure}[h]
	\centering
	\includegraphics[width=\linewidth]{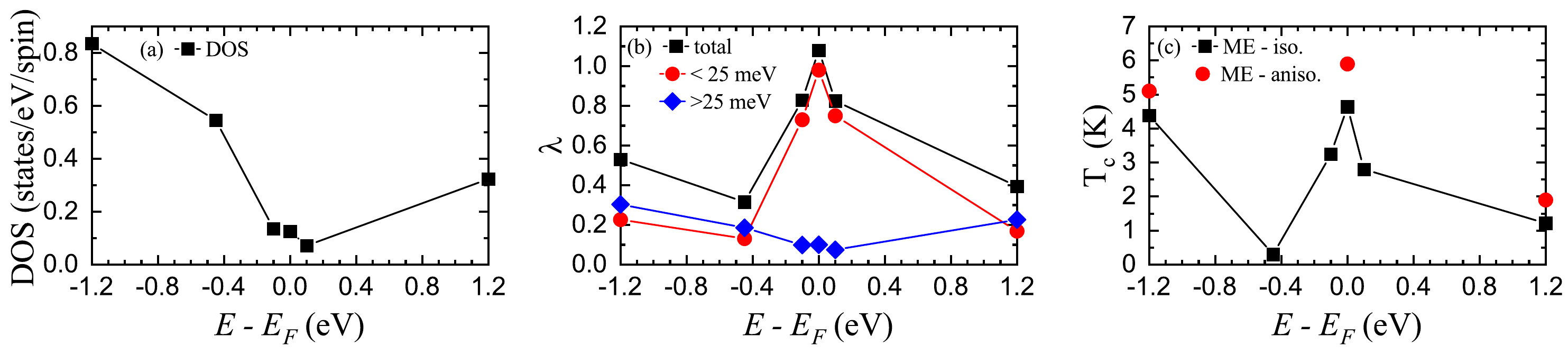}
	\caption{\label{figS19-dos-lambda-tc-SnS2-p30} Variations in (a) DOS at $E_{\rm F}$, (b) $\lambda$, and (c) $T_{\rm c}$ as a function of a rigid shift of the Fermi level for the $H$1 structure in the $\s1$ unit cell of SnS$_2$ at 30~GPa.  In (b), squares represent the total $\lambda$, while circles and rhombuses represent the contribution of the low- and high-energy modes. In (c), circles and squares represent the $T_{\rm c}$ obtained from the numerical solutions of the anisotropic and isotropic ME equations.}
\end{figure}

\begin{figure}[h]
	\centering
	\includegraphics[width=\linewidth]{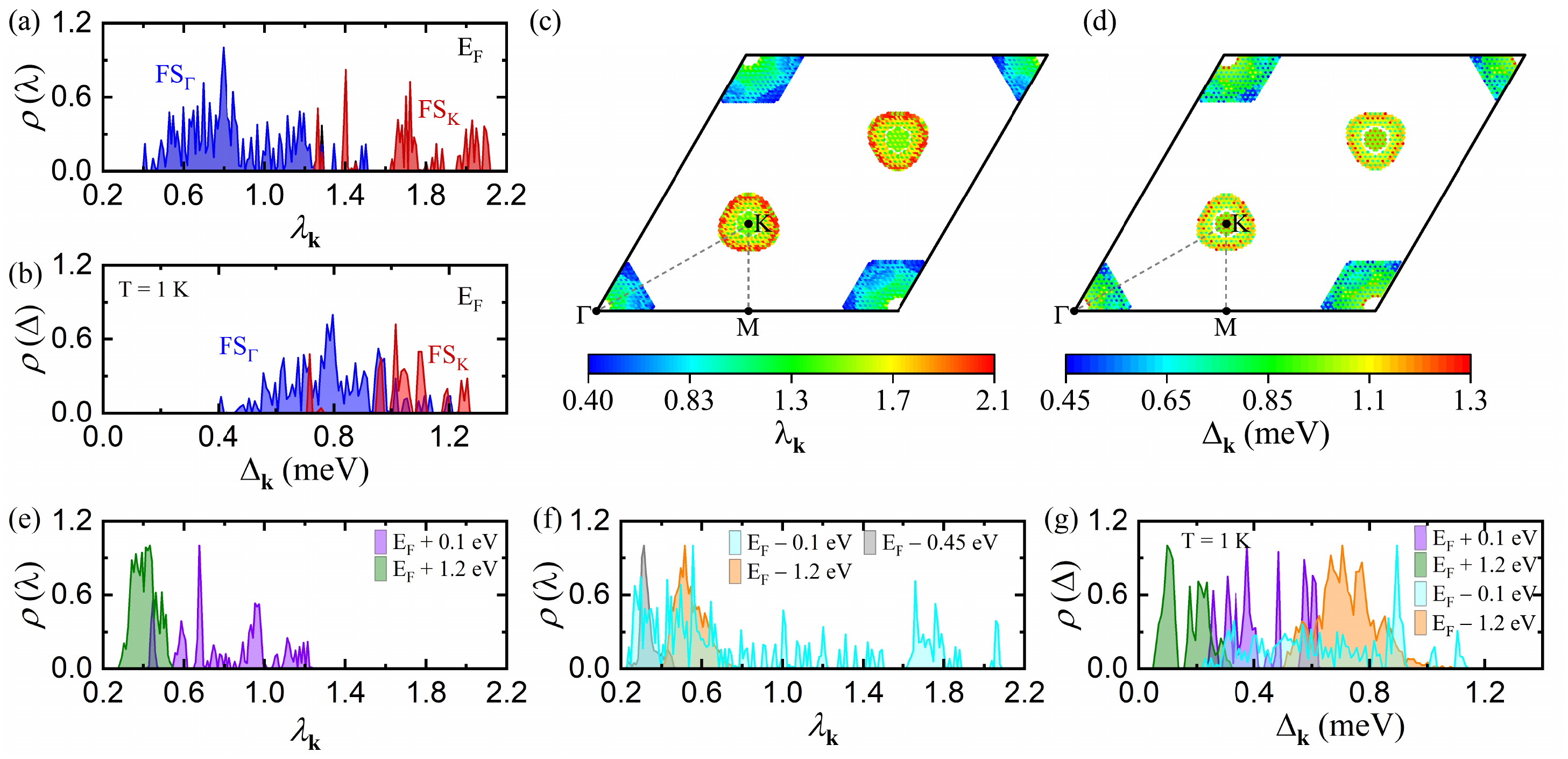}
	\caption{\label{figS20-SnS2_h1-supercond-p30} Calculated superconducting properties for the $H$1 structure in the $\s1$ unit cell of SnS$_2$ at 30~GPa. Energy distribution of the (a) e-ph coupling strength $\lambda_{\bf k}$ and (b) superconducting gap $\Delta_{\bf k}$; color coded by FS sheets: $\Gamma$-centered holelike pocket ${\rm FS}_\Gamma$ (blue), and $K$-centered electronlike pocket ${\rm FS}_K$ (red). Momentum-resolved (c) e-ph coupling strength $\lambda_{\bf k}$ and (d) superconducting gap $\Delta_{\bf k}$ on the FS (top-view). Energy distribution of the (e) e-ph coupling strength $\lambda_{\bf k}$, and (f)-(g) superconducting gap $\Delta_{\bf k}$ at $T = 1$~K at various rigid shifts of the Fermi level with respect to the original data as shown in Fig.~\ref{figS17-SnS2_h1-band-dos-p30}(a)-(b).}
\end{figure}

\begin{figure}[h]
	\centering
	\includegraphics[width=\linewidth]{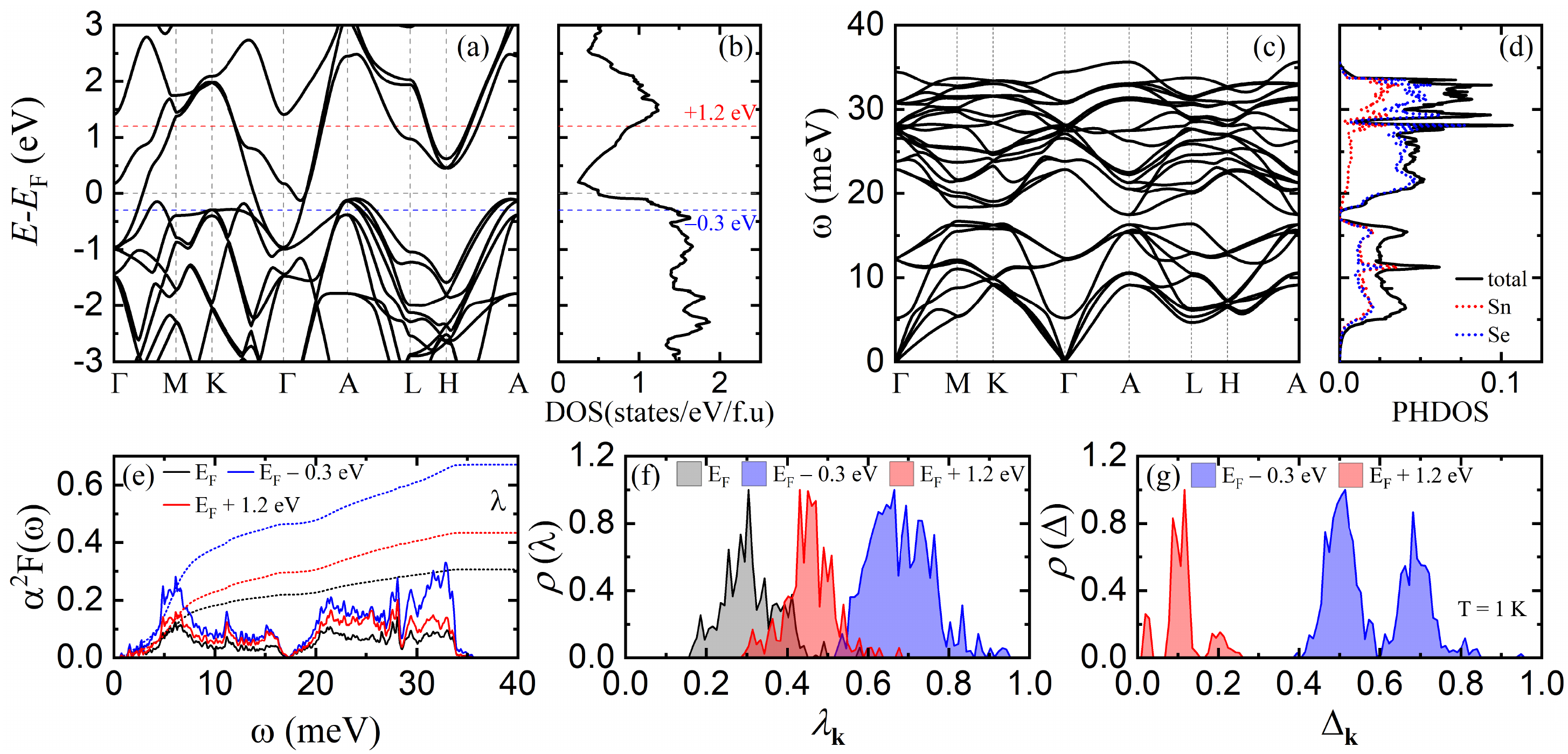}
	\caption{\label{figS21-SnSe2_h21-band-dos-ph-a2f-p30} Calculated (a) band structure, (b) DOS, (c) phonon dispersion, (d) PHDOS, (e) Eliashberg spectral function $\alpha^2F(\omega)$ and e-ph coupling strength $\lambda(\omega)$, (f) energy distribution of the e-ph coupling strength $\lambda_{\bf k}$, and (g) energy distribution of the superconducting gap $\Delta_{\bf k}$ at $T = 1$ K for the $H$2-1 structure in the $\sq3$ supercell of SnSe$_2$ at 30 GPa. In (e)-(g), black lines show the quantities at the Fermi level, while red and blue lines correspond to the rigid shifts in the Fermi level indicated in (a) and (b). Note that the DOS and PHDOS are plotted per formula unit.}
\end{figure}

\begin{figure}[h]
	\centering
	\includegraphics[width=\linewidth]{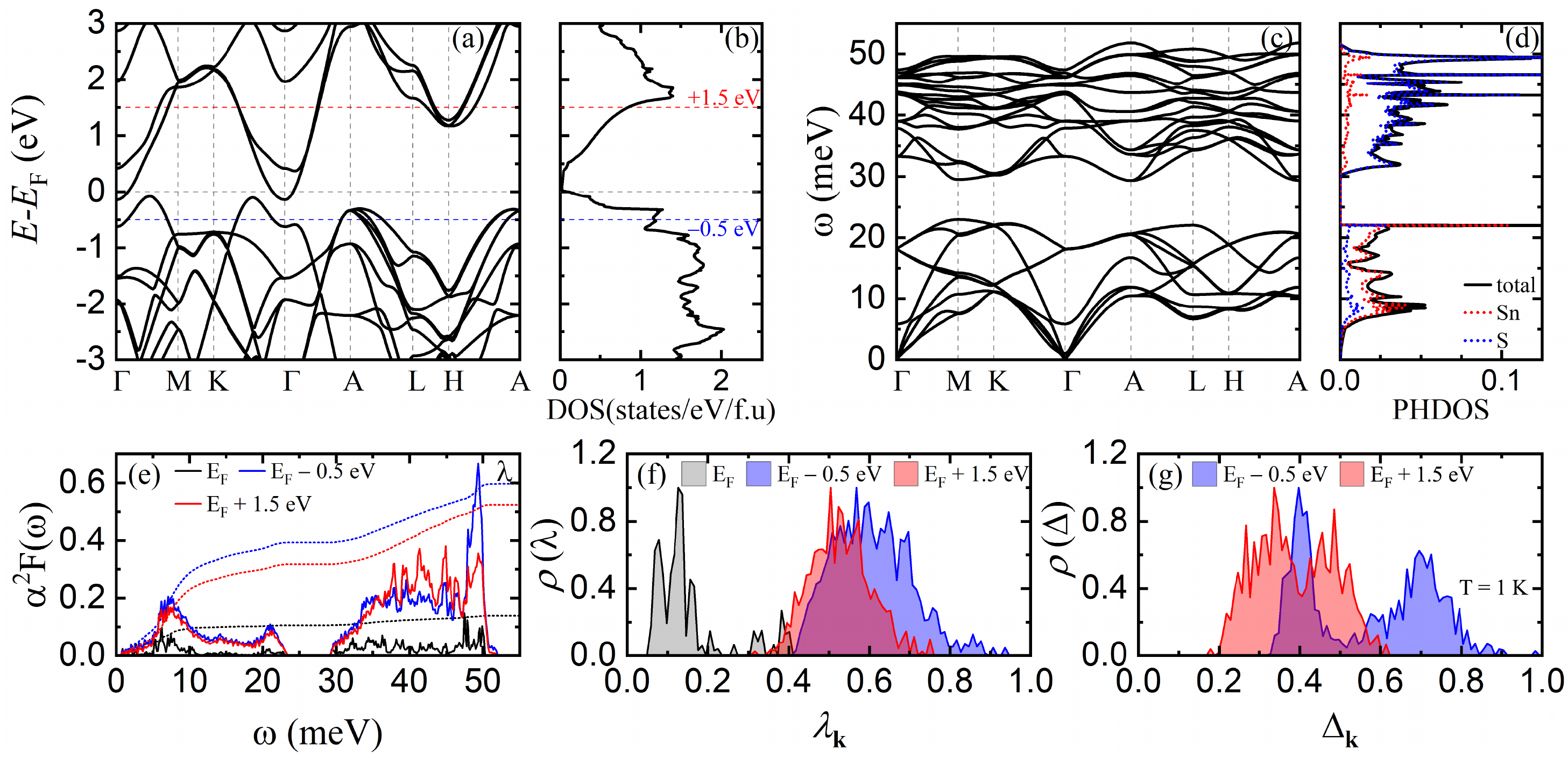}
	\caption{\label{figS22-SnS2_h21-band-dos-ph-a2f-p30} Calculated (a) band structure, (b) DOS, (c) phonon dispersion, (d) PHDOS, (e) Eliashberg spectral function $\alpha^2F(\omega)$ and e-ph coupling strength $\lambda(\omega)$, (f) energy distribution of the e-ph coupling strength $\lambda_{\bf k}$, and (g) energy distribution of the superconducting gap $\Delta_{\bf k}$ at $T = 1$ K for the $H$2-1 structure in the $\sq3$ supercell of SnS$_2$ at 30 GPa. In (e)-(g), black lines show the quantities at the Fermi level, while red and blue lines correspond to the rigid shifts in the Fermi level indicated in (a) and (b). Note that the DOS and PHDOS are plotted per formula unit.}
\end{figure}

\begin{figure}[h]
	\centering
	\includegraphics[width=\linewidth]{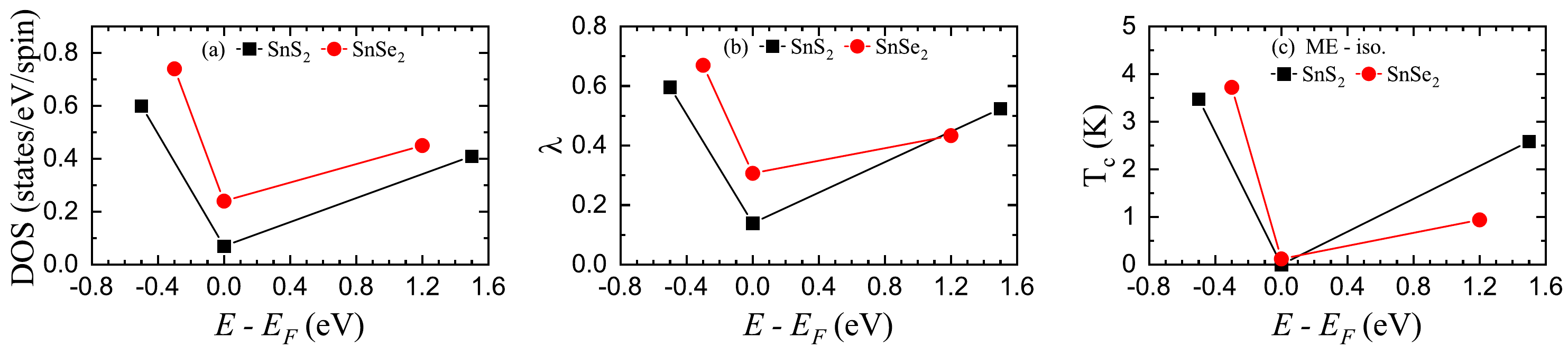}
	\caption{\label{figS23-H21-SnS2-SnSe2-dos-lambda-tc-p30} Variations in (a) DOS at $E_{\rm F}$, (b) $\lambda$, and (c) $T_{\rm c}$ as a function of a rigid shift of the Fermi level for the $H$2-1 structure in the $\sq3$ supercell of SnSe$_2$ (red lines and symbols) and SnS$_2$ (black lines and symbols) at 30~GPa. In (c), the $T_{\rm c}$ is obtained from the numerical solutions of the isotropic ME equations.}
\end{figure}

\FloatBarrier